\documentclass[%
 aip,
 amsmath,amssymb,
 reprint,%
]{revtex4-1}

\usepackage{graphicx}
\usepackage{dcolumn}
\usepackage{bm}

\usepackage[utf8]{inputenc}
\usepackage[T1]{fontenc}
\usepackage{mathptmx}
\usepackage{etoolbox}

\makeatletter
\def\@email#1#2{%
 \endgroup
 \patchcmd{\titleblock@produce}
  {\frontmatter@RRAPformat}
  {\frontmatter@RRAPformat{\produce@RRAP{*#1\href{mailto:#2}{#2}}}\frontmatter@RRAPformat}
  {}{}
}%

\usepackage{tikz}
\usetikzlibrary{arrows.meta}
\usetikzlibrary{positioning}

\usepackage{mathpazo} 
\usepackage{upgreek}

\renewcommand{\thetable}{S\arabic{table}}
\renewcommand{\thefigure}{S\arabic{figure}}
\renewcommand{\theequation}{S\arabic{equation}}

\usepackage{multirow}
\usepackage[colorlinks=true, linkcolor=blue, urlcolor=blue, citecolor=blue, anchorcolor=blue]{hyperref}

\usepackage{graphicx}
\usepackage{dcolumn}
\usepackage{bm}
\usepackage{helvet}

\newcommand{\unpolarizedSymbol}{
\begin{tikzpicture}[baseline=0.006mm]
    \draw[line width=0.4mm] (0,0.1) circle[radius=0.2cm];
    \draw[line width=0.15mm][<->, >=latex] (-0.2,0.1) -- (0.2,0.1);
    \draw[<->, >=latex] [line width=0.15mm] (0,-0.1) -- (0,0.3);
\end{tikzpicture}
}

\makeatother
\begin{document}

\preprint{AIP/123-QED}

\title[Optical communications through highly scattering channels using the coherence-rank]{Optical communications through highly scattering channels using the coherence-rank\\}
\author{Mitchell Harling}
\affiliation{PROBE Lab, School of Engineering, Brown University, Providence, RI 02912, USA}

\author{Chandler Stevenson}%
\affiliation{PROBE Lab, School of Engineering, Brown University, Providence, RI 02912, USA}

\author{Kimani C. Toussaint, Jr.}
\affiliation{PROBE Lab, School of Engineering, Brown University, Providence, RI 02912, USA}
\affiliation{Brown University Center for Digital Health, Providence, RI 02912, USA.}
 
\author{Ayman F. Abouraddy}
\affiliation{%
CREOL, The College of Optics \& Photonics, University of Central Florida, Orlando, Florida 32816, USA
}%
\email{raddy@creol.ucf.edu.}
\email{kimani\_toussaint@brown.edu.}
\date{\today}

\maketitle

\section*{abstract}
In optical communications, logical bits are encoded in physical degrees-of-freedom (DoFs) of the electromagnetic field. Consequently, optical scattering in a communications channel compromises the information transfer. In the worst-case-scenario, bit-to-bit stochastically varying scattering that couples the DoFs to each other -- including even unused DoFs -- can decouple the transmitter and receiver when relying on conventional physical encoding schemes, and preclude the utilization of adaptive techniques as a counter-measure. Here we show that partially coherent optical fields help circumvent the worst-case-scenario of rapidly varying, strong optical scattering, even when the channel is rendered informationally opaque for conventional approaches. Using a channel in which the spatial and polarization DoFs are relevant, we encode the logical bits in the unitarily invariant coherence rank (the number of non-zero eigenvalues of the field coherence matrix) and prepare our partially coherent fields with maximal entropies of 0, 1, 1.585, and 2 bits for rank-1, rank-2, rank-3, and rank-4 coherence matrices, respectively. This scheme demonstrates scattering-immune optical communications with 100 $\%$ fidelity. These results unveil an unexpected utility for partially coherent light in optical communications through challenging environments.

\section{Introduction}
\begin{figure}[t!]
\includegraphics[width=8.5cm]{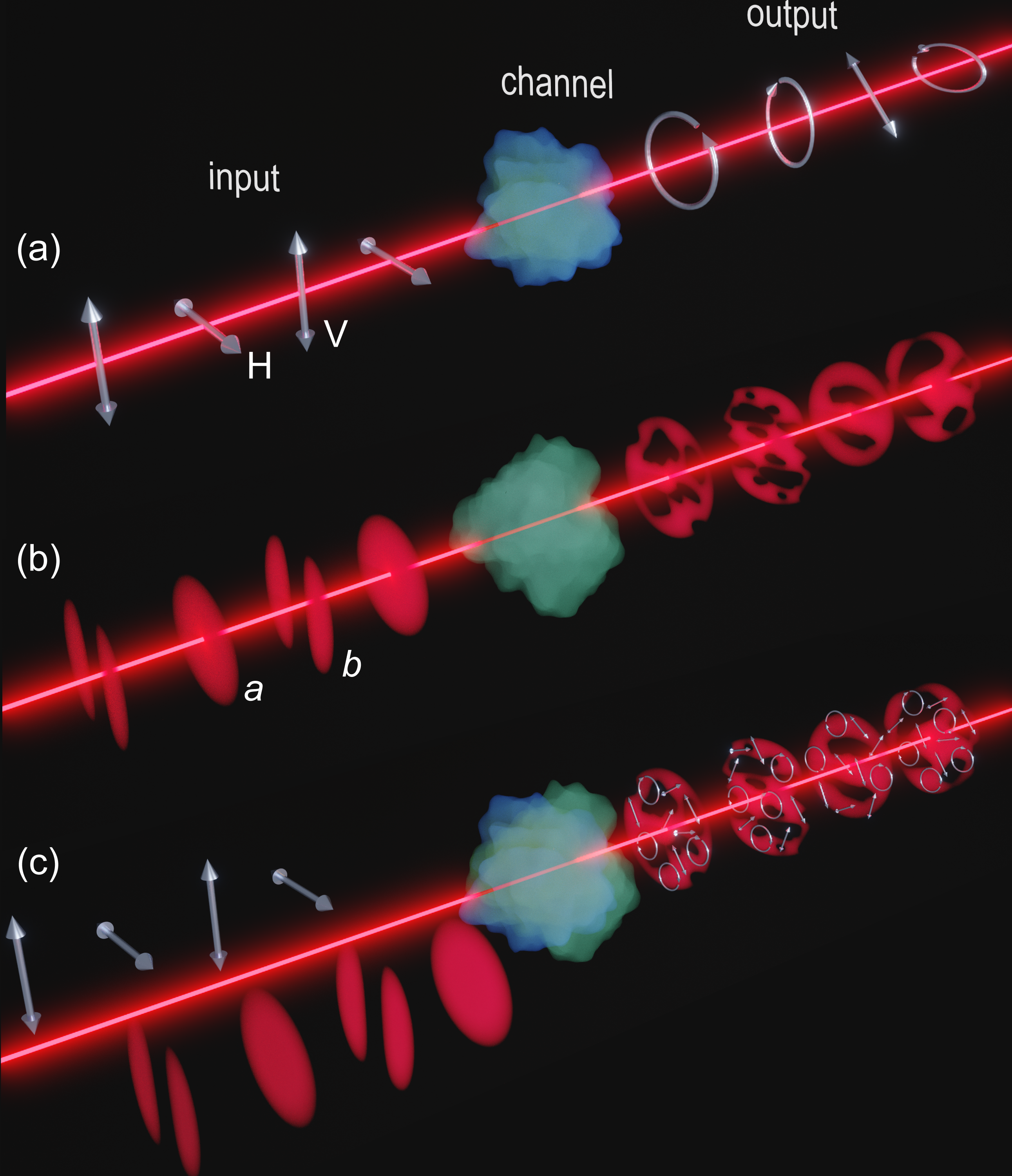}
\caption{\textbf{Optical communications over a strongly scattering channel.} (a) The logical bits are mapped to pure, linear polarization states (e.g., $0\rightarrow$H and $1\rightarrow$V), and the channel scrambles the polarization state. (b) The logical bits are mapped to coherent spatial modes (e.g., $0\rightarrow a$ and $1\rightarrow b$), and the channel scrambles the spatial modes. (c) The logical bits are mapped either to pure linear polarization states as in (a) or to coherent spatial modes as in (b), but the channel scrambles both DoFs (the used and unused DoFs) and couples them randomly to each other.}
\label{Fig:Concept}
\end{figure}

Whether for communications or computation, information must be encoded in physical states \cite{Landauer91PT}, which opens up the possibility of information corruption in a channel that varies randomly in time. To optimize their distinguishability after traversing an optical communications channel that adds noise and introduces scattering, logical bits are typically encoded in orthogonal physical states -- whether in different amplitudes (e.g., on/off keying) \cite{Agrawal2010Book}, polarization states \cite{Ursin07NPhys} [Fig.~\ref{Fig:Concept}(a)], or spatial modes \cite{Li14AOP,Willner15AOP} [Fig.~\ref{Fig:Concept}(b)]. A scattering channel (e.g., a multimode optical fiber) randomly couples these physical states (e.g., the fiber modes) \cite{Ho14JLT,Cao23AOP}. A variety of strategies have been explored to combat such scattering. Adaptive techniques assume the channel variations occur at a slower rate than the transmitted data stream, so that infrequent channel probing helps inform the encoding process \cite{Tyson91Book,Duffner09Book,Ren14Optica,Carpenter15NPhot,Defienne18PRL,Valencia20NPhys,Zhou21NC}, and are thus not useful when the physical channel changes bit-to-bit. Another strategy embeds the logical space in a subspace of a larger-dimensional physical space, which is dynamically decoupled from the environment, thereby mitigating channel-induced errors -- as developed for error correction in quantum communications and computation \cite{Walton03PRL}. This approach requires significantly larger physical resources, it has not been tested in strongly scattering environments or rapidly varying channels, and it has not yet been adapted for communications with classical light. Another approach relies on devising physical states of the optical field that are themselves immune to scattering \cite{Nelson14JOSAA,Zhao15OL,Cox16OE,Mphuthi18JOSAA,Zhu21NC}. The status of these proposals with respect to their efficacy in combating optical scattering remains unclear.

An intriguing possibility in the optical domain is to make use of the untapped potential of partially coherent light \cite{Born99Book,Wolf07Book,Goodman15Book}. Applications of partially coherent light have relied to date on their unique \textit{physical} characteristics; for example, through measurements of the coherence time or coherence width \cite{Baleine05PRL}, or speckle statistics \cite{Goodman07Book,Kondakci17SR,Han23PRL,Bender23Optica}. However, it has long been noted that partially coherent optical fields possess more free parameters than their coherent-field counterparts of the same dimensionality \cite{Waller12NP}. A coherent field in a physical space spanned by $N$ modes requires $2N-2$ real parameters for its determination (grows \textit{linearly} with $N$), while its partially coherent counterpart in the same space requires $N^{2}-1$ (grows \textit{quadratically} with $N$). As such, the synthesis and analysis of partially coherent fields is more demanding, but may offer a potential enhancement in the information-carrying capacity as suggested theoretically in the context of optical communications through a multimode fiber \cite{Nardi22OL}.

Another feature of realistic optical communications channels is that they usually deleteriously impact multiple DoFs of the field \cite{Vitullo17PRL,Xiong2018} [Fig.~\ref{Fig:Concept}(c)], whether used or unused in encoding the information, which further corrupts the transmission. For concreteness, we consider here a model optical channel in which the spatial and polarization DoFs are relevant. When the spatial DoF comprises two modes, the most general coherence matrix $\mathbf{G}$ associated with the polarization and spatial DoFs has dimensions $4\times4$ (Supplementary Material Section 1). We have recently studied this class of coherence matrices \cite{Abouraddy17OE,Okoro17Optica} and examined the limits of entropy transfer between the two DoFs under global unitary transformations spanning both DoFs (henceforth, `unitaries' for brevity) \cite{harling2022reversible,Harling23JO}. We have found that the `coherence rank' of the field -- the number of non-zero eigenvalues of $\mathbf{G}$ -- serves as a convenient classifier for categorizing partially coherent fields with respect to their behavior under unitaries \cite{Harling24PRA,Harling24PRA2}. 

We consider here channels that represent the worst-case scattering scenario for optical communications (without complete loss of signal): (1) the channel strongly scatters both the polarization and spatial DoFs, so that using either DoF for communications fails; (2) the scattering strongly couples the two DoFs (potentially with maximal inter-DoF crosstalk), resulting in spatially dependent polarization changes or polarization-dependent spatial changes; and (3) the channel changes unpredictably from bit to bit, with no long-range correlation in time. We pose here the following question: Is it possible to devise physical states of the optical field that are immune to such rapidly varying, strong scattering, and thus reliably achieve error-free communications?

Here, we show that partially coherent optical states of this dual-DoF field can be constructed to transmit information that is immune to the worst-case-scenario of optical scattering in a communications channel. We first confirm experimentally that scattering in this channel is sufficient to fully scramble transmission when information is encoded in a single DoF (polarization here). We then establish that encoding the logical bits in the coherence rank of the field encompassing both the polarization and spatial DoFs is immune to both intra-DoF and inter-DoF scattering. The coherence rank is thus a stable structural field parameter by virtue of its invariance under any unitary transformation coupling the two DoFs, in addition to losses impacting the DoFs as long as they are rank-preserving. We construct here partially coherent, two-point vector fields of different rank, and transmit the field through strongly scattering channels that vary erratically from bit to bit by encoding the logical bits in the maximum-entropy field configuration associated with each rank. At the channel output, we tomographically reconstruct the coherence matrix from projective measurements, extract the eigenvalues, and estimate the coherence rank, from which we confirm the immunity of the rank to strong scattering. The results of this proof-of-principle experiment can be useful in optical communications through challenging environments, and sensing and imaging through turbid media.

\section{Models for the scattering channel and encoding schemes}

We first briefly describe a conventional scenario of encoding information in a single DoF of the field. We consider exploiting a single binary optical DoF, such as polarization, that is separable from the other DoFs [Fig.~\ref{Fig:Concept}(a)]. Bits 0 and 1 are encoded in two states represented by the $2\times2$ polarization coherence matrices $\mathbf{G}_{0}$ and $\mathbf{G}_{1}$, respectively, which are Hermitian, unity-trace, and positive semi-definite, $\scriptsize \mathbf{G}_{\mathrm{p}}\!=\!\left(\begin{array}{cc}G^{\mathrm{HH}}&G^{\mathrm{HV}}\\G^{\mathrm{VH}}&G^{\mathrm{VV}}\end{array}\right)$, where $G^{jk}\!=\!\langle E^{j}E^{k*}\rangle$, $E^{j}$ is a scalar component of the electric field, $j$ and $k$ correspond to either the horizontal (H) or vertical (V) linear polarization states with respect to some fixed direction, and $\langle\cdot\rangle$ is the ensemble average \cite{Wolf07Book}.

We make the following assumptions about the optical communications channel: (1) the channel impacts only the polarization DoF; (2) losses in the channel are not polarization-sensitive; (3) the channel is reversible and can be represented -- after factoring out the losses -- by a unitary transformation $\hat{U}_{\mathrm{p}}$; (4) $\hat{U}_{\mathrm{p}}$ is selected randomly and uniformly from the complete set of possible unitary transformations (strong scattering); (5) $\hat{U}_{\mathrm{p}}$ changes bit-to-bit in time; and (6) $\hat{U}_{\mathrm{p}}$ at time $t$ is statistically uncorrelated with $\hat{U}_{\mathrm{p}}$ at any other time. This is an extreme model, and most realistic channels relax assumptions (4) through (6): the channel typically changes at a slower rate than the data, and $\hat{U}_{\mathrm{p}}$ is \textit{not} selected uniformly from the set of all transformations (weak scattering). Nevertheless, retaining these extreme assumptions highlights the unique advantage of coherence-rank communications. One can establish an analogous communications channel using a pair of spatial modes [Fig.~\ref{Fig:Concept}(b)]. Here $\scriptsize \mathbf{G}_{\mathrm{s}}\!=\!\left(\begin{array}{cc}G_{aa}&G_{ab}\\G_{ba}&G_{bb}\end{array}\right)$ is the $2\times2$ spatial coherence matrix in analogy with $\mathbf{G}_{\mathrm{p}}$, with $a$ and $b$ identifying two spatial modes \cite{Halder21OL}.

\begin{figure*}[t!]
\includegraphics[width=17cm]{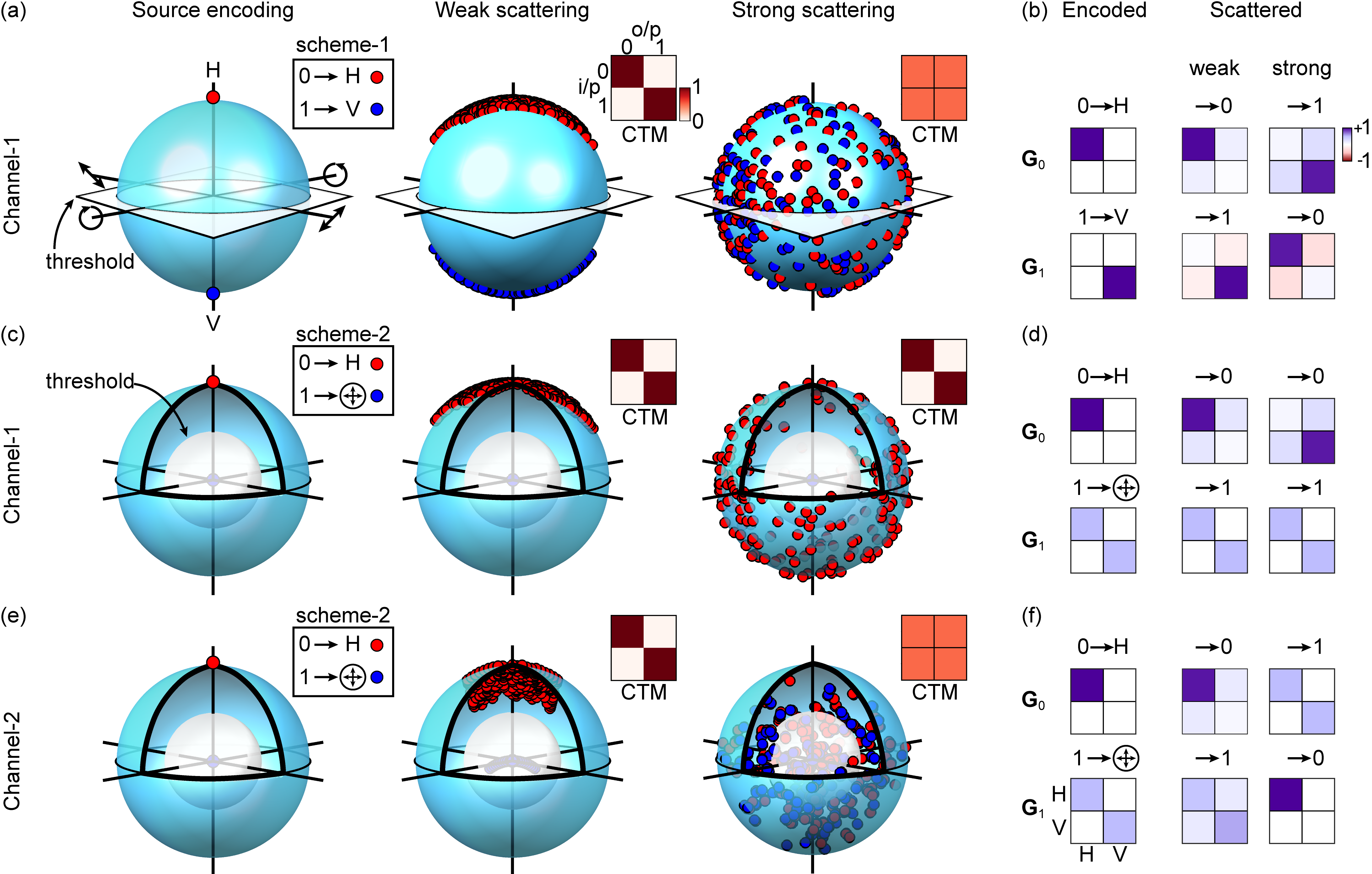}
\caption{\textbf{Optical communications using the polarization DoF through scattering channels represented on the unity-radius Poincar{\'e} sphere (PS).} 
(a) Encoding scheme-1 is shown on the left where logical bits are encoded in orthogonal polarizations, horizontal (H) and vertical (V), which correspond to opposite poles on the PS. The decision threshold is the equator plane midway between H and V, and the inset represents the encoding scheme (here $0\rightarrow\mathrm{H}$ and $1\rightarrow\mathrm{V}$). The middle and right panels show the output represented on the PS after traversing a channel that weakly and strongly scatters polarization (Ch-1), respectively, along with the associated the cross-talk matrix (CTM) as insets. (b) The single-DoF polarization coherence matrices ($\mathbf{G}_{\mathrm{p}}$) associated with the encoded and scattered (weakly and strongly) bits for encoding scheme-1 through Ch-1. Although weakly scattered coherence matrices can still be decoded correctly, the strongly scattered coherence matrices cannot. (c) Same as (a) after utilizing encoding scheme-2, where the logical bits are encoded in polarized and unpolarized field states, corresponding to a pole ($P=1$) and the center ($P=0$) of the PS. The decision threshold is the spherical surface of radius $\tfrac{1}{2}$. (d) Same as (b); however, encoding scheme-2 succeeds where encoding scheme-1 fails. Any scattered coherence matrix can still be decoded even in the strong scattering limit. (e) Same as (c) utilizing encoding scheme-2, but to traverse a scattering channel that couples the polarization DoF to an unused spatial DoF (Ch-2). This channel can be polarizing (i.e., increasing the degree of polarization of the output field relative to the input field), or depolarizing (i.e., decreasing the degree of polarization). Therefore the point representing the field can move anywhere across the PS volume. (f) Same as (d); however the strongly scattered coherence matrices can no longer be decoded.}
\label{Fig:Solution}
\end{figure*}

The customary encoding scheme maps the bits 0 and 1 to orthogonal polarization states, say H and V [Fig.~\ref{Fig:Solution}(a)], which correspond to the two poles of the unity-radius Poincar{\'e} sphere (PS), in which case $\scriptsize \mathbf{G}_{0}\!=\!\left(\begin{array}{cc}1&0\\0&0\end{array}\right)$ and $\scriptsize \mathbf{G}_{1}\!=\!\left(\begin{array}{cc}0&0\\0&1\end{array}\right)$ [Fig.~\ref{Fig:Solution}(d)], and we set the threshold for deciding the output bit at the equator plane. This encoding scheme is susceptible to a scattering channel $\hat{U}_{\mathrm{p}}$ that couples the polarization components and moves the point representing the field on the PS surface. For weak scattering, the points corresponding to $\mathbf{G}_{0}$ and $\mathbf{G}_{1}$ migrate away from the PS poles, but rarely cross the equator. The cross-talk matrix representing the coupling between the input and output logical states is diagonal [Fig.~\ref{Fig:Solution}(a), weak scattering], as required for a low bit-error-rate (BER). In presence of strong scattering, $\mathbf{G}_{0}$ and $\mathbf{G}_{1}$ are equally likely after the channel to migrate to any position on the PS, the cross-talk matrix is consequently flat [Fig.~\ref{Fig:Solution}(a), strong scattering], and the transmitter and receiver are effectively decoupled (BER$\sim50\%$). We consider examples of scattered coherence matrices in Fig.~\ref{Fig:Solution}(b) starting from the coherence matrices $\mathbf{G}_{0}$ and $\mathbf{G}_{1}$. For weak scattering, the resulting coherence matrices still resemble the encoding coherence matrices and can thus be successfully decoded. In contrast, for strong scattering this is not always the case. Indeed, in the example shown in Fig.~\ref{Fig:Solution}(b), the channel basically swaps H and V, thus eliminating any chance of correctly decoding the signal. If the scattering channel does indeed change rapidly, then adaptive optics techniques are precluded as a remedy.

The impact of such severe scattering can be nevertheless thwarted by employing a different encoding strategy that makes use of optical coherence. Consider mapping 0 to pure H polarization (the PS north pole), $\scriptsize \mathbf{G}_{0}\!=\!\left(\begin{array}{cc}1&0\\0&0\end{array}\right)$, and mapping 1 to unpolarized light (the PS center), $\scriptsize \mathbf{G}_{1}\!=\!\tfrac{1}{2}\left(\begin{array}{cc}1&0\\0&1\end{array}\right)$; see Fig.~\ref{Fig:Solution}(c,d). The two fields differ in their degree of polarization $P$, which corresponds to the distance from the PS center to the point representing the field, and is defined as $P\!=\!|\lambda_{1}-\lambda_{2}|$, where $\lambda_{1}$ and $\lambda_{2}$ are the eigenvalues of $\mathbf{G}_{\mathrm{p}}$; $P\!=\!1$ for $\mathbf{G}_{0}$ and $P\!=\!0$ for $\mathbf{G}_{1}$ \cite{Brosseau06PO,Wolf07Book,Abouraddy17OE}. The strongly scattering channel considered above does not change $P$, so we set the decision threshold at the spherical surface $P\!=\!\tfrac{1}{2}$. Weak scattering results in $\mathbf{G}_{0}$ migrating on the PS surface in the vicinity of the north pole, and strong scattering spreads the migration across the entire PS surface. Nevertheless, because $\mathbf{G}_{1}$ is unaffected by either scattering regime, the decision threshold at $P\!=\!\tfrac{1}{2}$ distinguishes $\mathbf{G}_{0}$ and $\mathbf{G}_{1}$, and the cross-talk matrix is diagonal even in presence of weak or strong scattering.
The right column of Fig.~\ref{Fig:Solution}(d) shows that the eigenvalues of $\mathbf{G}_{0}$ and $\mathbf{G}_{1}$ can be recovered to determine $P$. This encoding scheme is thus impervious to the extreme channel model adopted here. Moreover, \textit{any} point on the PS surface would do for bit 0. To the best of our knowledge, this scheme has yet to be implemented for communications. 

\begin{figure*}[t!]
\centering
\includegraphics[width=10cm]{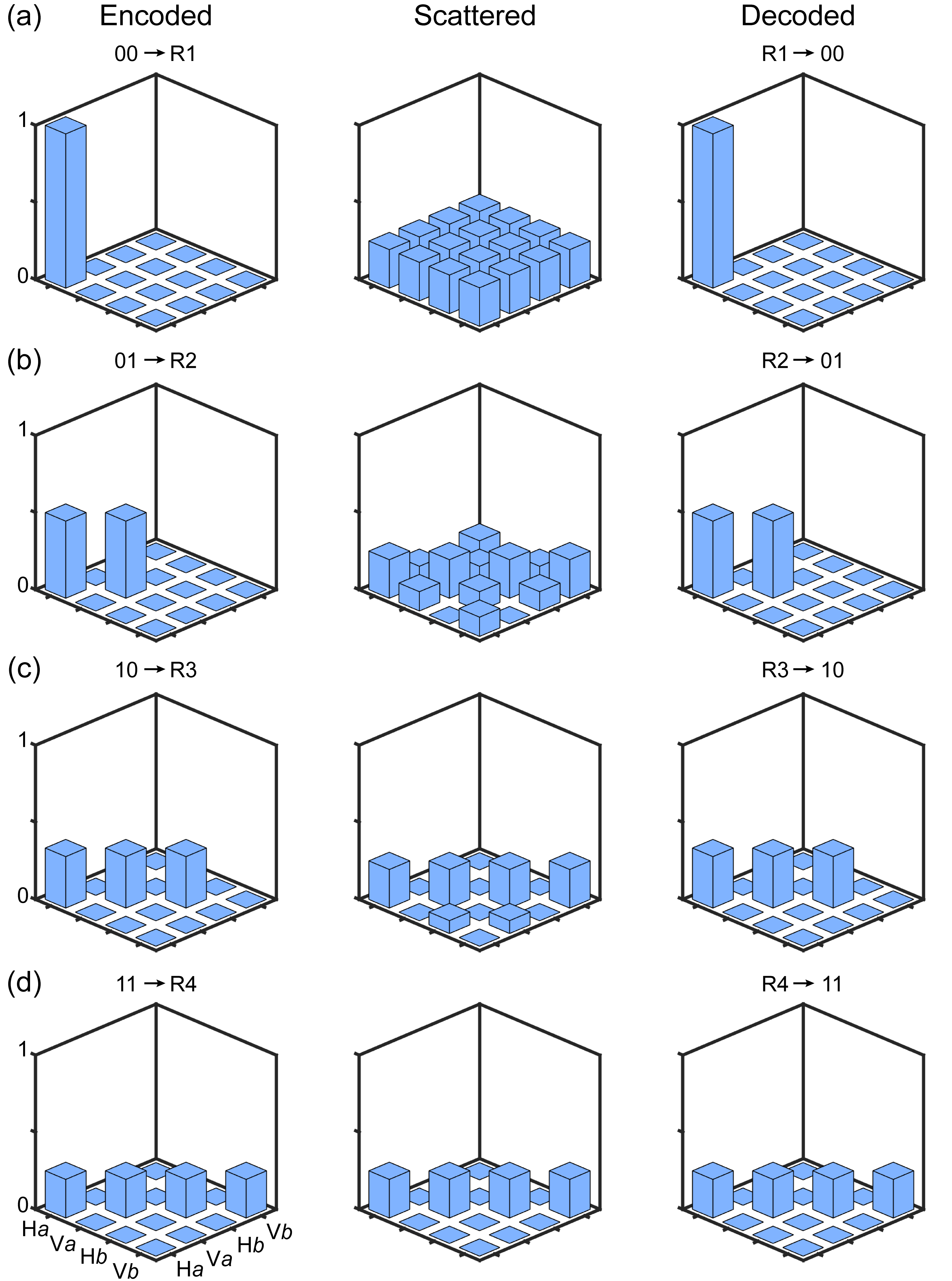}
\caption{\textbf{Optical coherence-rank communication through unitarily scattering channels.} The left column shows the multi-DoF coherence matrices ($\mathbf{G}$); the middle column shows an instant of the multi-DoF coherence matrices after scattering in a channel represented by a unitary transformation representative of Ch-2 [Fig.~\ref{Fig:Solution}(e,f)]; and the coherence matrices in the right column are obtained from their counterparts in the middle column after diagonalization. We make use here of encoding scheme-3, where (a) a rank-1 field (R1) represents bits 00, (b) a rank-2 field (R2) represents bits 01, (c) a rank-3 field (R3) represents bits 10, and (d) a rank-4 field (R4) represents bits 11. For each rank, the coherence matrix selected corresponds to a maximum-entropy field. The maximum-entropy rank-4 field is the identity matrix, which is invariant under any unitary transformation. All the matrices displayed are the real parts of $\mathbf{G}$.
}
\label{Fig:matrices}
\end{figure*}

\begin{figure*}[t!]
\centering
\includegraphics[width=15cm]{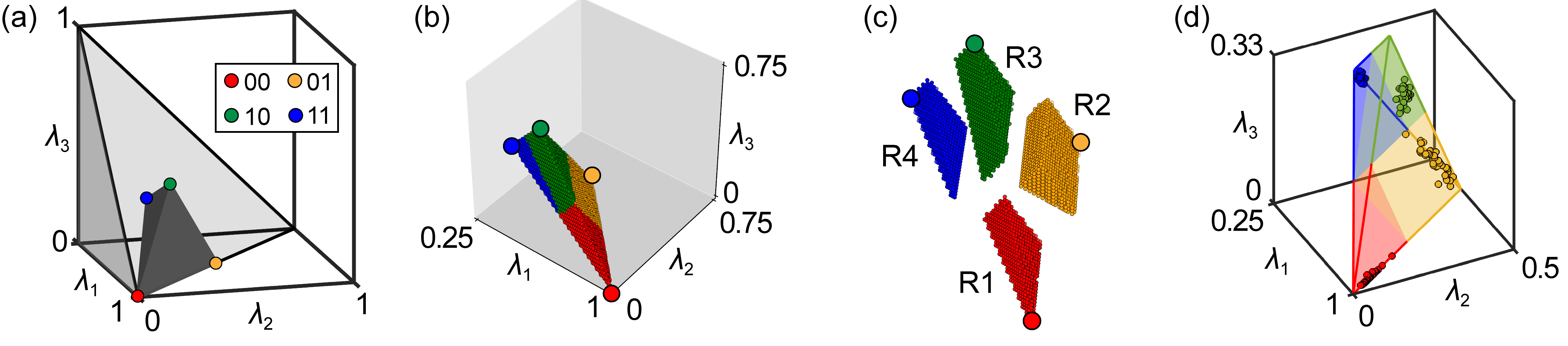}
\caption{\textbf{Optical communications via the coherence rank through a scattering optical channel.} (a) Representation of the diagonal coherence matrix $\mathbf{G}\!=\!\mathrm{diag}\{\lambda_{1},\lambda_{2},\lambda_{3},\lambda_{4}\}$ in a reduced 3D space spanned by $\lambda_{1}$, $\lambda_{2}$, and $\lambda_{3}$, relying on $\lambda_{4}\!=\!1-(\lambda_{1}+\lambda_{2}+\lambda_{3})$. The pyramid volume corresponds to the projected space of all diagonalized coherence matrices. The smaller identified volume corresponds to coherence matrices with eigenvalues arranged with descending values $\lambda_{1}\geq\lambda_{2}\geq\lambda_{3}\geq\lambda_{4}$. The legend identifies the two-bit encoding scheme, in which the bits are associated with the states at the vertices of the highlighted sub-volume, which correspond to the maximum-entropy fields for each rank. (b) Magnified view of the irreducible volume identified in (a). (c) Exploded view of the volume in (b). The surfaces separating the sub-volumes are the decision thresholds for the logical alphabet code. (d) Same as (b) but with the experimental outcomes for the reconstructed $\mathbf{G}$ (encoding scheme-3 in Fig.~\ref{Fig:MeasurementsCh2}) plotted, which cluster around the vertices.}
\label{Fig:CoherenceRank}
\end{figure*}

We offer a rationale for the efficacy of partially coherent light here that paves the way to understanding coherence-rank communications. The parameters involved in defining $\mathbf{G}_{\mathrm{p}}$ may be divided heuristically into `radial' parameters that determine the radius from the PS center of a spherical surface on which the point representing the field lies (determined by the eigenvalues of $\mathbf{G}_{\mathrm{p}}$), and `angular' parameters that determine the position of the point on this spherical surface. The scattering channel impacts the angular parameters, whereas the radial parameters are invariant. Coherent (polarized) optical fields, which lie on the PS outer surface, are dependent on the angular parameters, and are inevitably affected by scattering. Instead, relying on the radial parameter associated with the degree of polarization $P$ renders the communications scheme independent of $\hat{U}_{\mathrm{p}}$. This comes at the price of eschewing the use of the angular parameters. Rather than exploit the additional real parameters required to identify the coherence matrix to increase the information-carrying capacity as proposed in ref \cite{Nardi22OL}, here we sacrifice these extra parameters to help circumvent the impact of scattering in the channel.

However, this encoding scheme is \textit{not} immune to de-polarizing channels that change $P$ [Fig.~\ref{Fig:Solution}(e)], which arises when the scattering channel couples polarization to an unused DoF, say the spatial DoF \cite{Vitullo17PRL}. 
We retain the above-listed assumptions for the channel, but add a further stipulation: the channel (represented by a $4\times4$ unitary transformation $\hat{U}$ \cite{Abouraddy17OE}) couples the polarization and spatial DoFs arbitrarily bit-to-bit. Despite the unitarity of $\hat{U}$, $P$ is no longer invariant: the channel may reduce $P$ (potentially resulting in the conversion $\mathbf{G}_{0}\rightarrow\mathbf{G}_{1}$) or increase $P$ ($\mathbf{G}_{1}\rightarrow\mathbf{G}_{0}$) [Fig.~\ref{Fig:Solution}(f), strong scattering]. Although the total entropy is invariant, entropy can be exchanged between the DoFs \cite{Okoro17Optica,harling2022reversible}, thereby rendering a polarized field unpolarized, or an unpolarized field polarized ($\mathbf{G}_{0}\leftrightarrow\mathbf{G}_{1}$).

To describe this scenario quantitatively, we first extend the description of the field to a $4\times4$ coherence matrix $\mathbf{G}$ that encompasses both DoFs \cite{Gori06OL,Abouraddy17OE,Okoro17Optica,harling2022reversible,Harling23JO,Harling24PRA,Harling24PRA2} (Supplementary Material Section~1). We assume the field is initially in a separable state, $\mathbf{G}\!=\!\mathbf{G}_{\mathrm{p}}\otimes\mathbf{G}_{\mathrm{s}}$. The information is encoded in the polarization DoF, and the detector reconstructs only the reduced polarization coherence matrix $\mathbf{G}_{\mathrm{p}}^{\mathrm{red.}}$ after tracing out the unused spatial DoF \cite{Kagalwala13NP}. When the channel does not couple the DoFs, then $\mathbf{G}_{\mathrm{p}}^{\mathrm{red.}}\!=\!\mathbf{G}_{\mathrm{p}}$, and the scenario in Fig.~\ref{Fig:Solution}(a-d) follows [the channel in Fig.~\ref{Fig:Concept}(a)]. However, when the channel couples the DoFs, then $\mathbf{G}_{\mathrm{p}}^{\mathrm{red.}}\!\neq\!\mathbf{G}_{\mathrm{p}}$; see Fig.~\ref{Fig:Solution}(e,f) [the channel in Fig.~\ref{Fig:Concept}(c)]. If the channel causes both polarization and spatial scattering (intra-DoF scattering), and couples the polarization and spatial DoFs (inter-DoF scattering), then the point representing any input polarization state moves around the entire PS volume after traversing such a channel. Although the decision threshold at $P\!=\!\tfrac{1}{2}$ can distinguish $\mathbf{G}_{0}$ and $\mathbf{G}_{1}$ in presence of weak scattering (diagonal cross-talk matrix), the two points corresponding to $\mathbf{G}_{0}$ and $\mathbf{G}_{1}$ spread across the entire PS volume in presence of strong scattering, and thus cannot be distinguished by the decision threshold at $P\!=\!\tfrac{1}{2}$ (flat cross-talk matrix). See Fig.~\ref{Fig:Solution}(f) for coherence matrices in this situation---the original eigenvalues cannot be recovered. This physical-channel model is the worst-case scenario from the point of view of polarization scattering. Communications across this channel via polarization is not viable.

Nevertheless, utilizing partially coherent fields in the full space encompassing both DoFs can help overcome this apparently insurmountable obstacle. 
By exploiting both DoFs, we encode pairs of bits in the rank of $\mathbf{G}$. The space of all coherence matrices $\mathbf{G}$ can be represented on a Poincar{\'e} hypersphere in 16 dimensions. Any point on the Poincar{\'e} hypersphere for two binary DoFs is determined by 16 real parameters: 12 `angular' parameters, and 4 `radial' parameters corresponding to the eigenvalues of $\mathbf{G}$ after diagonalization, whereupon $\mathbf{G}\!=\!\mathrm{diag}\{\lambda_{1},\lambda_{2},\lambda_{3},\lambda_{4}\}$, $\lambda_{j}$ are the non-negative eigenvalues of $\mathbf{G}$, $\sum_{j=1}^{4}\lambda_{j}\!=\!1$, and all off-diagonal elements are set to zero. The eigenvalues are the weights of the so-called coherent modes \cite{Wolf82JOSA}. Diagonalization eliminates all the angular parameters and retains only the radial parameters. For a single binary DoF, we have two radial parameters (the eigenvalues $\lambda_{1}$ and $\lambda_{2}$), which are reduced to one after normalizing the trace. For two binary DoFs, we have 4 radial parameters, which are reduced to 3 after normalization. Whereas rank-1 (R1) fields are uniquely defined by the coherence matrix $\mathbf{G}\!=\!\mathrm{diag}\{1,0,0,0\}$ in the diagonal representation [Fig.~\ref{Fig:matrices}(a)], fields of higher rank afford broad flexibility in selecting a representative member associated with the rank.
For example, R2 fields $\mathbf{G}\!=\!\mathrm{diag}\{\lambda_{1},\lambda_{2},0,0\}$ with $\lambda_{1}+\lambda_{2}\!=\!1$ make up a one-parameter family of fields that have the same rank but differ in entropy $S\!=\!-\lambda_{1}\mathrm{log}_{2}\lambda_{1}-\lambda_{2}\mathrm{log}_{2}\lambda_{2}$ \cite{Harling24PRA,Harling24PRA2} [Fig.~\ref{Fig:matrices}(b)]. This flexibility extends to two- and three-parameter families of fields for R3 [Fig.~\ref{Fig:matrices}(c)] and R4 [Fig.~\ref{Fig:matrices}(d)]. We employ the following encoding scheme: $00\rightarrow$R1, $01\rightarrow$R2, $10\rightarrow$R3, and $11\rightarrow$R4 (here R1 refers to rank-1 fields, etc.). 

We have selected here the coherence matrix corresponding to the maximum entropy for each rank: for R1 we choose $\mathbf{G}\!=\!\mathrm{diag}\{1,0,0,0\}$ ($S\!=\!0$), for R2 we choose $\mathbf{G}\!=\!\tfrac{1}{2}\mathrm{diag}\{1,1,0,0\}$ ($S\!=\!1$~bit), for R3 we choose
$\mathbf{G}\!=\!\tfrac{1}{3}\mathrm{diag}\{1,1,1,0\}$ ($S\!\approx\!1.585$~bits), and for $\mathbf{G}\!=\!\tfrac{1}{4}\mathrm{diag}\{1,1,1,1\}$ ($S\!=\!2$~bits), which can be seen in the "encoded" coherence matrices in the left column of Fig.~\ref{Fig:matrices}. While the angular parameters of the coherence matrices are affected by scattering in Ch-2 (middle column of Fig.~\ref{Fig:matrices}) whereby the structure of the coherence matrix can change significantly. Nevertheless, the radial parameters (the eigenvalues of $\mathbf{G}$) can always be recovered by diagonalization, even in the presence of inter-DoF scattering, to decode the rank (right column of Fig.~\ref{Fig:matrices}).

By restricting the Poincar{\'e} hypersphere to a 4D space spanned by $\{\lambda_{1},\lambda_{2},\lambda_{3},\lambda_{4}\}$, each $\mathbf{G}$ corresponds to a point on a plane embedded in that space. Projecting this 4D plane onto a restricted 3D space spanned by $\{\lambda_{1},\lambda_{2},\lambda_{3}\}$ results in the pyramid volume shown in Fig.~\ref{Fig:CoherenceRank}(a), with any $\mathbf{G}$ corresponding to a point in this volume. When ordering the eigenvalues in descending order ($\lambda_{1}\!\geq\!\lambda_{2}\!\geq\!\lambda_{3}\!\geq\!\lambda_{4}$), the volume is reduced to that shown in Fig.~\ref{Fig:CoherenceRank}(a,b). The coherence matrices that we have chosen for coherence-rank encoding shown in the left column of Fig.~\ref{Fig:matrices} correspond to the vertices of the reduced volume in Fig.~\ref{Fig:CoherenceRank}(a,b).
Defining a Euclidean metric in this space of eigenvalues (Appendix: Methods), this reduced volume is sub-divided into volume segments corresponding to states of the optical fields that are `closest' to each of the vertices under a Euclidean metric [the color-coded sub-volumes in Fig.~\ref{Fig:CoherenceRank}(b,c)]. The surfaces separating these volume segments are the decision thresholds for detection. The depolarizing channel in the strong scattering regime corresponding to Fig.~\ref{Fig:Solution}(c) (third column) has no impact on the selected vertices, so that the resulting $4\times4$ cross-talk matrix is diagonal. Coherence-rank communications is thus viable even through this worst-case scattering channel with no need to resort to adaptive techniques. Note that utilizing the coherent modes for communications instead (the modal basis after diagonalizing $\mathbf{G}$ \cite{Wolf82JOSA}) remains susceptible to this scattering channel.

\section{Experiment}

\begin{figure*}[t!]
\centering
\includegraphics[width=17cm]{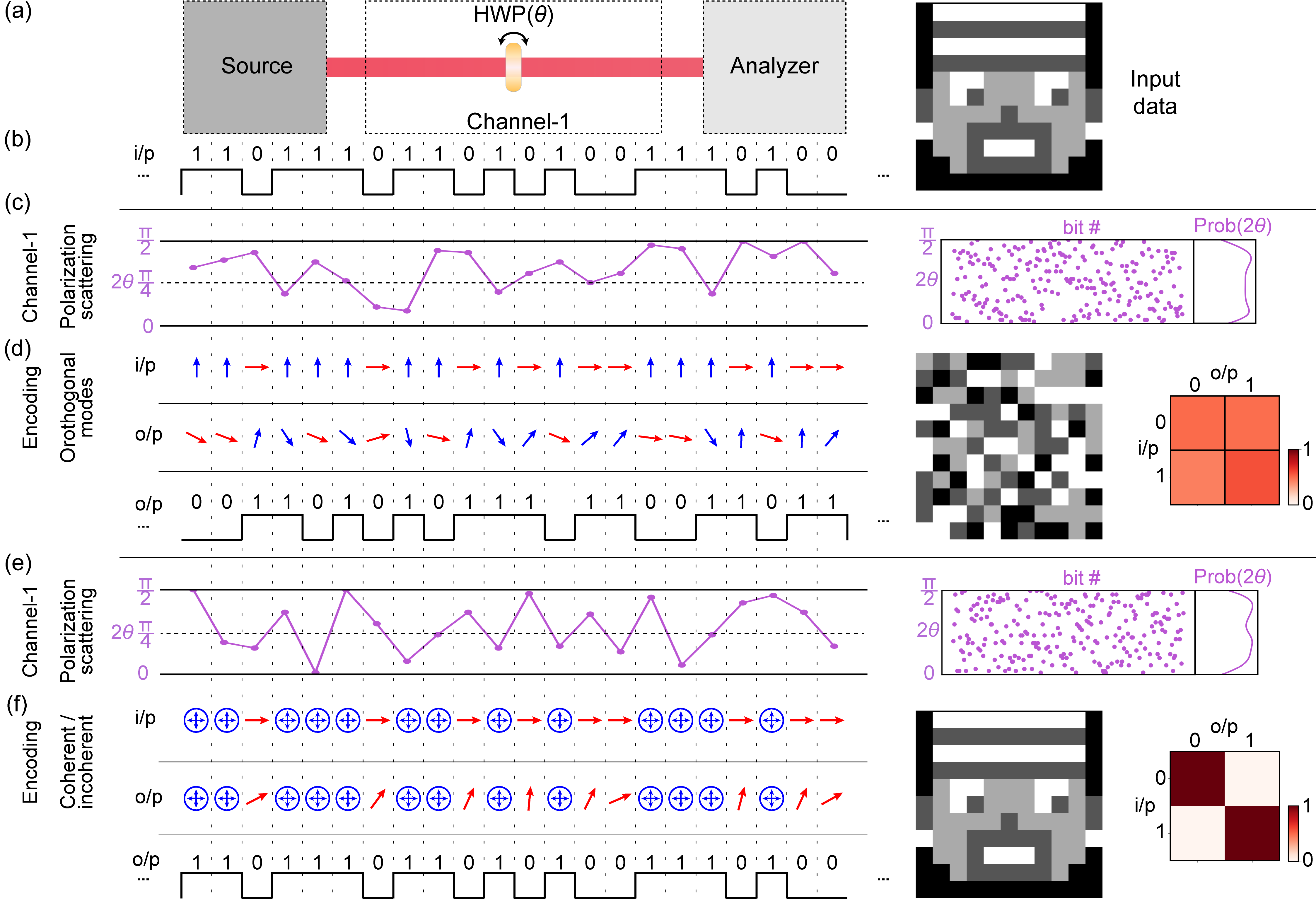}
\caption{\textbf{Measurement results for optical communications through a polarization scattering channel Ch-1.} (a) Schematic of the setup for Ch-1. (b) Portion of the data stream corresponding to the image on the right. (c,d) Polarization encoding scheme-1, where the logical bits are mapped to two orthogonal linear polarization states: $0\rightarrow$H and $1\rightarrow$V [Fig.~\ref{Fig:Solution}(a)]. (c) Settings for the HWP angle $\theta$ in the channel. The probability distribution of $\theta$ is plotted on the right. (d) Input and output polarization states corresponding to the data stream. The measured cross-talk matrix is plotted on the right, along with reconstructed image. (e,f) Same as (c,d) for polarization encoding scheme-2, where the logical bits are mapped to polarized and unpolarized fields, $0\rightarrow$H and $1\rightarrow$ \protect\unpolarizedSymbol, respectively.}
\label{Fig:MeasurementsCh1}
\end{figure*}

To test the hypotheses outlined above, we prepare a data stream corresponding to an image [$11\times11$~pixels, 4~gray-scale levels each, for a total of 242 bits; Fig.~\ref{Fig:MeasurementsCh1}(a), right panel], and send the data stream [Fig.~\ref{Fig:MeasurementsCh1}(b)] through two model channels. The first channel (Ch-1) corresponds to an unknown polarization transformation in the strong scattering regime [Fig.~\ref{Fig:Solution}(a,b), third column], realized with a half-wave plate (HWP) rotated an angle $\theta$ with respect to a fixed axis, $\scriptsize \hat{U}(\theta)\!=\!\left(\begin{array}{cc}\cos2\theta&\sin2\theta\\\sin2\theta&-\cos2\theta\end{array}\right)$; see Fig.~\ref{Fig:MeasurementsCh1}(a) and Supplementary Material Section~3. We take the worst-case-scenario where $\theta$ changes randomly bit-to-bit with a probability density function for $\theta$ that is uniform over a the span $[0,45^{\circ}]$; Fig.~\ref{Fig:MeasurementsCh1}(c). The correlation length for $\theta$ is 1~bit (verified in Supplementary Material Section~3). We show in Fig.~\ref{Fig:MeasurementsCh1} the results for two encoding schemes (Appendix: Methods and Supplementary Material Section~2) through Ch-1. In encoding scheme-1 [Fig.~\ref{Fig:MeasurementsCh1}(d)], the data is encoded in linear polarization states, $0\rightarrow$H and $1\rightarrow$V. Here the optical source is an unpolarized LED that is spatially filtered, and the polarization is selected via a linear polarizer whose orientation is appropriately switched (Appendix: Methods and Supplementary Material Figure~S2). We maintain the spatial profile of the modes throughout the setup with collimating lenses, so that we can ignore the propagation-induced coherence changes captured in the van~Cittert-Zernike theorem \cite{Born99Book,Wolf07Book}. The coherence matrix for the output field is reconstructed using optical coherence matrix tomography (OCmT) following the protocol in Ref.~\cite{Abouraddy14OL,Kagalwala15SR} (Supplementary Material Section~5). The tomographic measurement setup as well as the source encoding are fully automated. Note that we here exploit the partial coherence of the LED, in contrast to LED-based LiFi communications that relies instead on amplitude modulation \cite{Dimitrov15Book}. 

The channel transforms the polarization states H and V into new linearly polarized states, resulting in a measured BER $\sim50\%$ and flat cross-talk matrix, indicating equal coupling between any input and output states [Fig.~\ref{Fig:MeasurementsCh1}(d)]. Consequently, the reconstructed image at the receiver is corrupted as expected from such a strongly scattering channel. In encoding scheme-2 (Appendix: Methods and Supplementary Material Figure~S2), we make use of polarized and unpolarized states, $0\rightarrow$H ($P\!=\!1$) and $1\rightarrow$unpolarized light ($P\!=\!0$), respectively, to encode the same data stream and traverse Ch-1 [Fig.~\ref{Fig:MeasurementsCh1}(e)]. We make use of the same spatially filtered unpolarized LED as the source, but now insert or remove a linear polarizer to produce polarized and unpolarized light [Fig.~\ref{Fig:MeasurementsCh1}(f) and Supplementary Material Figure~S2]. Because $P$ is invariant (Ch-1 is rank-preserving in the polarization space), this encoding scheme is found to be impervious to Ch-1 despite the randomizing HWP, The resulting cross-talk matrix is diagonal, and the image is faithfully reconstructed [Fig.~\ref{Fig:MeasurementsCh1}(f)]. Simulations for communications through Ch-1 with encoding scheme-1 and encoding scheme-2 are provided in Supplementary Material Figure~S5. 

\begin{figure*}[t!]
\centering
\includegraphics[width=17cm]{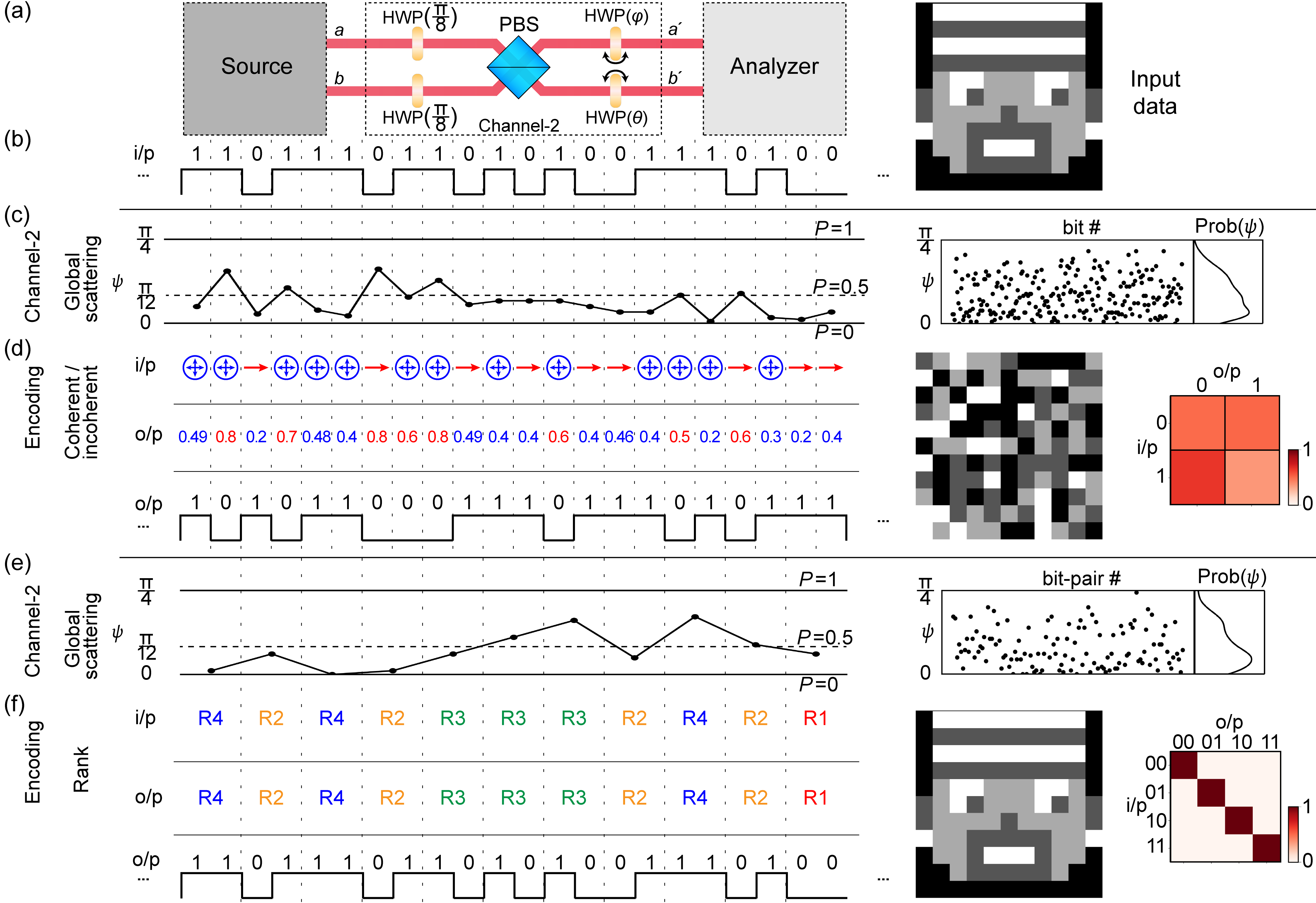}
\caption{\textbf{Measurement results for optical communications through Ch-2 that couples the polarization and spatial DoFs.} (a) Schematic of the setup for Ch-2. (b) Portion of the data stream corresponding to the image on the right [same as in Fig.~\ref{Fig:MeasurementsCh1}(b)]. (c,d) Polarization encoding scheme-2 from Fig.~\ref{Fig:MeasurementsCh1}(e,f): $0\rightarrow$H and $1\rightarrow$\protect\unpolarizedSymbol. (c) Settings for $\psi\!=\!|\theta-\varphi|$, where $\theta$ and $\varphi$ are the HWP angles in the channel, and the probability distribution of $\psi$ is plotted on the right. (d) Input and output polarization states corresponding to the data stream. The measured cross-talk matrix is plotted on the right, along with the reconstructed image. (e,f) Same as (c,d) for encoding scheme-3, coherence-rank communications. Pairs of logical bits are encoded in the coherence rank: $00\rightarrow$R1, $01\rightarrow$R2, $10\rightarrow$R3, and $11\rightarrow$R4. (f) The measured input and output coherence ranks reconstructed tomographically via OCmT.}
\label{Fig:MeasurementsCh2}
\end{figure*}

We now move on to a second scattering optical channel (Ch-2) that offers a significant challenge for optical communications [Fig.~\ref{Fig:MeasurementsCh2}(a)]. Here, both the polarization and spatial DoFs are relevant, where the spatial DoF comprises a pair of spatial modes, identified as $a$ and $b$, which correspond to two mutually incoherent points selected from the LED field (Appendix: Methods and Supplementary Material Figure~S3). A variety of randomly varying effects take place in the channel simultaneously: the polarization is rotated, the spatial DoF is modified, and the polarization and spatial DoFs are coupled to each other. We design the channel model in Fig.~\ref{Fig:MeasurementsCh2}(a) to be rank-preserving in the full $4\times4$ space spanned by polarization and spatial modes, but it is not rank-preserving in the restricted polarization space (after tracing out the spatial modes). In Ch-2, polarized light at the input may emerge unpolarized at the output, and unpolarized light can emerge polarized as a result of entropy swapping between the DoFs \cite{Okoro17Optica,harling2022reversible,Harling23JO}. Consequently, Ch-2 disrupts encoding scheme-1 and encoding scheme-2 that rely solely on the polarization DoF. Nevertheless, coherence-rank communications (encoding scheme-3) remains feasible because the rank in the full $4\times4$ space is preserved despite the random entropy swapping between the two DoFs.

We first examine the impact of Ch-2 on encoding scheme-2 that was used above successfully in Ch-1, and utilized solely the polarization DoF ($0\rightarrow$H and $1\rightarrow$unpolarized light). We make use of the same data stream in Ch-2 [Fig.~\ref{Fig:MeasurementsCh2}(b)] that was utilized in Fig.~\ref{Fig:MeasurementsCh1} for Ch-1. The optical field is initially separable with respect to the two DoFs, $\mathbf{G}\!=\!\mathbf{G}_{\mathrm{p}}\otimes\mathbf{G}_{\mathrm{s}}$, and the spatial DoF comprises a single mode $a$ (spatially coherent field) after blocking the second mode $b$ at the input (but any other spatially coherent field would do). Two HWP settings ($\theta$ and $\varphi$) are varied randomly over independent, uniform probability distributions (Supplementary Material Figure~S6), and the probability distribution of their absolute difference $\psi=|\varphi-\theta|$ is plotted in Fig.~\ref{Fig:MeasurementsCh2}(c). After the channel, we measure the polarization DoF only and trace over the spatial DoF, which corresponds to the \textit{reduced} coherence matrix $\mathbf{G}_{\mathrm{p}}^{\mathrm{red.}}\!=\!\mathrm{Tr}_{\mathrm{s}}\{\mathbf{G}\}$. Because Ch-2 couples the two DoFs, the initially separable state becomes non-separable \cite{Kagalwala13NP,Abouraddy17OE}, and consequently $\mathbf{G}_{\mathrm{p}}^{\mathrm{red.}}\!\neq\!\mathbf{G}_{\mathrm{p}}$. Indeed, for strong scattering, $\mathbf{G}_{\mathrm{p}}^{\mathrm{red.}}$ is altogether decoupled from $\mathbf{G}_{\mathrm{p}}$, and the measured cross-talk matrix is now flat -- in contrast to the outcome for the same encoding scheme in Ch-1 (where the cross-talk matrix is diagonal); Fig.~\ref{Fig:MeasurementsCh2}(d).

The second encoding scheme employed in Ch-2 makes use of the coherence rank, encoding scheme-3 [Fig.~\ref{Fig:MeasurementsCh2}(e)]. Here we encode pairs of bits of the data stream in partially coherent optical fields of different rank, making use of the maximum-entropy field representative of each rank in the diagonal representation. We thus have $00\rightarrow$~R1, $01\rightarrow$~R2, $10\rightarrow$~R3, and $11\rightarrow$~R4. The points associated with the $4\times4$ coherence matrices generated at the source (Appendix: Methods and Supplementary Material Figure~S3) correspond to vertices of the highlighted, restricted sub-volume in Fig.~\ref{Fig:CoherenceRank}(a,b). In reconstructing the $4\times4$ coherence matrices $\mathbf{G}$ at the source and after Ch-2, we employ a maximum likelihood estimation technique \cite{James01PRA} to ensure that $\mathbf{G}$ satisfies the criteria for a coherence matrix (in particular, to exclude coherence matrices that violate positivity). Ch-2 substantially modifies $\mathbf{G}$ produced at the source by coupling the two DoFs to each other, thereby converting the representation of $\mathbf{G}$ from diagonal to non-diagonal. At the channel output, we reconstruct $\mathbf{G}$ tomographically via joint spatial-polarization Stokes-parameter measurements (Supplementary Material Section~5), extract the eigenvalues of $\mathbf{G}$, and estimate its rank. Measurement errors at the receiver in reconstructing $\mathbf{G}$ (unrelated to the randomly varying Ch-2) lead to a deviation in the point representing $\mathbf{G}$ from the ideal points at the vertices of the restricted sub-volume in Fig.~\ref{Fig:CoherenceRank}(b,d). Utilizing a Euclidean metric in the $\{\lambda_{1},\lambda_{2},\lambda_{3},\lambda_{4}\}$-space, we estimate which sub-volume segment the is the optimal decision for the received signal (Appendix: Methods). We plot the measured $\mathbf{G}$ reconstructed at the Ch-2 output in Fig.~\ref{Fig:MeasurementsCh2}(d), where we note that the output points cluster around the vertices but do not cross the boundary surfaces between the sub-volume segments, which correspond to the decision thresholds under a Euclidean metric. Consequently, the measured $4\times4$ cross-talk matrix is now diagonal, and the image is reconstructed faithfully [Fig.~\ref{Fig:MeasurementsCh2}(f)]. In other words, coherence-rank communications in encoding scheme-3 survives the worst-case scattering scenario realized in Ch-2. Simulations of communications through Ch-2 with encoding scheme-2 and encoding scheme-3 are provided in Supplementary Material Figure~S7.

\section{Discussion}

In addition to the possibility of establishing scattering-immune optical communications in presence of the worst-case-scenario of scattering, coherence-rank communications offers further advantages. First, this scheme solves the problem of frame-sharing; that is, the sender and receiver need not share the same polarization axes for defining the physical states of the field. A polarization misalignment between the sender and receiver corresponds to an additional unknown unitary transformation, which does not affect the coherence rank. Indeed, the sender may choose to vary the polarization basis bit-to-bit without impacting the receiver's ability to decode the information. Similarly, phase offsets between the spatial or polarization modes that may be incurred in the communications channel do not impact the coherence rank. Although overall losses are non-unitary, they do not impact this scheme because they are rank-preserving. Modal-dependent losses (barring modal extinction) are also rank-preserving, yet they may impact our scheme by potentially moving $\mathbf{G}$ across the boundaries of the domains in Fig.~\ref{Fig:Solution}(c,d), thereby leading to communications errors. Note that the concept of partial spatial coherence from a practical perspective does not extend to the entire electromagnetic spectrum \cite{Wolf07Book,SalehBook07}, where the size of the sources and detectors are usually on the order of the wavelength. It is thus not expected that this scheme is practical for radio or microwave frequencies.

One limitation of the coherence-rank communications scheme is that it fails when the rate of changes in the channel exceeds the data rate. A more significant challenge concerns the practical viability of realizing high-speed communications links utilizing partially coherent light. These concerns were recently allayed in Ref.~\cite{roques2024measuring} where it was shown that the general integrated-photonics approach known as `programmable photonics' \cite{Bogaerts20Nature} can be equally implemented with partially coherent light as with coherent light. If the polarization DoF is to be used, the scheme for scalar fields in Ref.~\cite{roques2024measuring} needs to be modified accordingly. Furthermore, the work in Ref.~\cite{roques2024measuring} points to the possibility for recent advances in deep learning to impact such schemes by reducing the number of measurements required given the larger number of free parameters defining the coherence matrix $\mathbf{G}$ with respect to a coherent field of the same dimension. This suggests the feasibility of high-speed optical communications with partially coherent light. Although our proposed coherence-rank communications scheme is mainly designed for challenging environments, it is nevertheless worth examining the limits on spectral utilization of the source bandwidth and its impact on the maximum data rate achievable, in addition to other sources of inter-symbol interference (e.g., multipath interference and dispersion). Finally, more work needs to be done to assess the impact of additive noise on the channel capacity, and thus extending the Hartley-Shannon theorem to coherence-rank communications, especially for high-rank values.

\section{Conclusion}
The proof-of-principle demonstration presented here is a critical step towards realizing scattering-immune optical communications. There are several directions in which this work can be immediately developed. First, although the three schemes shown in Fig.~\ref{Fig:Solution} make use of the polarization DoF, the same argument symmetrically applies to utilizing the spatial DoF for optical communications. Such a scheme can be particularly beneficial in light of recent rapid advances in spatial-mode multiplexing in multimode optical fibers. Relying on partially coherent light may thus provide a solution for the hurdle resulting from modal scattering at fiber bends and imperfections, which may also couple the spatial DoF to polarization. Rather than using two orthogonal modes to encode the information, one may utilize coherent and incoherent spatial fields, which is immune to scattering among these modes, but is thwarted by coupling to a second DoF such as polarization. Second, the coherence-matrix formulation can be readily extended to larger-dimensional DoFs of the optical field (equivalently, bigger encoding alphabets). For example, spatial modes in a multimode optical fiber, laser lines in a frequency comb, or time-bin communications. In general, for two DoFs represented in spaces of dimensions $N_{1}$ and $N_{2}$, the associated coherence matrix has dimensions $(N_{1}\times N_{2})\times(N_{1}\times N_{2})$, with $N_{1}\times N_{2}$ eigenvalues (potentially offering independent communications channels of that number). Third, a variety of optical channels can be investigated besides multimode optical fibers, including biological samples, underwater optical communication channels, the turbulent atmosphere, and turbid media.

\section*{Supplementary Material}
The supplementary material includes further details on the source preparation, encoding schemes, channels, and the tomographic analysis used to estimate the coherence rank.

\begin{acknowledgments}
This work was supported by the U.S. Office of Naval Research (ONR) under contracts N00014-17-1-2458 and N00014-20-1-2789.
\end{acknowledgments}

\section*{Data Availability Statement}
The data that support the findings of this study are available from the corresponding author upon reasonable request.

\section*{Author Contributions}
A.F.A. developed the concept on optical-rank communications. K.C.T. supervised the research. M.H. carried out the measurements, calculations, and analysis, with assistance from C.S., A.F.A., and K.C.T. All the authors contributed to the writing of the manuscript.

\appendix*
\section{Methods}

\noindent\textbf{Source preparation and encoding schemes.} The source is a light-emitting diode (LED) with a center wavelength of 625~nm, and a bandwidth (FWHM) of $\approx\!17$~nm (Supplementary Material Figure~S2a). A spectral bandpass filter (not shown in Supplementary Material Figure~S2) that has a center wavelength of 620~nm and a FWHM-bandwidth of 10~nm is used to improve the temporal coherence of the field by reducing the bandwidth. The field is restricted spatially via two vertically-oriented narrow slits to two spatial modes $a$ and $b$. The width of each slit is 100~$\mu$m, and they are separated by 23~mm (Supplementary Material Figure~S2a, inset). The separation distance is selected to be larger than the spatial coherence width of the LED. The field produced by the source in this configuration is the maximum-entropy rank-4 field described by the coherence matrix $\mathbf{G}_\mathrm{o}\!=\!\mathrm{diag}\{\tfrac{1}{4},\tfrac{1}{4},\tfrac{1}{4},\tfrac{1}{4}\}$; that is, the two spatial modes are mutually incoherent, and the field is unpolarized.

We made use of three different encoding schemes in the main text. To prepare the different source configurations used in the different encoding schemes in our optical communications experiments, we made use of the non-unitary operators listed below.

\textit{Encoding scheme-1.} The logical bits of 0 and 1 are encoded as H and V, respectively: $0\rightarrow$H and $1\rightarrow$V. We assume that the field is separable with respect to the spatial and polarization DoFs, and that the field is associated with mode $a$ (spatially coherent). The coherence matrices for these two field configurations (corresponding to bits 0 and 1) are:
\begin{eqnarray}
\mathbf{G}_{0}&=&\left(\begin{array}{cc}1&0\\0&0\end{array}\right)\otimes\left(\begin{array}{cc}1&0\\0&0\end{array}\right)=\mathrm{diag}\{1,0,0,0\},\nonumber\\
\mathbf{G}_{1}&=&\left(\begin{array}{cc}0&0\\0&1\end{array}\right)\otimes\left(\begin{array}{cc}1&0\\0&0\end{array}\right)=\mathrm{diag}\{0,1,0,0\},
\end{eqnarray}
respectively. These can be obtained from the source $\mathbf{G}_{\mathrm{o}}$ via the non-unitary transformations:
\begin{equation}
\hat{T}_{0}=\left(\begin{array}{cc}\hat{T}_\mathrm{H}&\mathbf{0}\\\mathbf{0}&\hat{T}_\mathrm{BB}\end{array}\right),\;\;\;\hat{T}_{1}=\left(\begin{array}{cc}\hat{T}_\mathrm{V}&\mathbf{0}\\\mathbf{0}&\hat{T}_\mathrm{BB}\end{array}\right),
\end{equation}
which yield the H and V polarized fields as required, respectively; $\hat{T}_\mathrm{BB}\!=\!\left(\begin{array}{cc}0&0\\0&0\end{array}\right) = \mathbf{0}$ (beam block),  $\hat{T}_\mathrm{H}\!=\!\left(\begin{array}{cc}1&0\\0&0\end{array}\right)$ (H-polarizer), and $\hat{T}_\mathrm{V}\!=\!\left(\begin{array}{cc}0&0\\0&1\end{array}\right)$ (V-polarizer),
see Supplementary Material Figure~S2b.

\textit{Encoding scheme-2.} The logical bits 0 and 1 are encoded in H and unpolarized fields, respectively: $0\rightarrow$H and $1\rightarrow$\protect\unpolarizedSymbol. Once again, we take the field to be separable with respect to the spatial and polarization DoFs, and the spatial DoF is associated with mode $a$ (spatially coherent). The coherence matrices for the field configurations corresponding to bits 0 and 1 are thus given by:
\begin{eqnarray}
\mathbf{G}_{0}&=&\left(\begin{array}{cc}1&0\\0&0\end{array}\right)\otimes\left(\begin{array}{cc}1&0\\0&0\end{array}\right)=\mathrm{diag}\{1,0,0,0\},\nonumber\\
\mathbf{G}_{1}&=&\frac{1}{2}\left(\begin{array}{cc}1&0\\0&1\end{array}\right)\otimes\left(\begin{array}{cc}1&0\\0&0\end{array}\right)=\frac{1}{2}\mathrm{diag}\{1,1,0,0\},
\end{eqnarray}
respectively. These can be obtained from the source $\mathbf{G}_{\mathrm{o}}$ via the non-unitary transformations:
\begin{equation}
\hat{T}_{0}=\left(\begin{array}{cc}\hat{T}_\mathrm{H}&\mathbf{0}\\\mathbf{0}&\hat{T}_\mathrm{BB}\end{array}\right),\;\;\;\hat{T}_{1}=\left(\begin{array}{cc}\mathbb{I}&\mathbf{0}\\\mathbf{0}&\hat{T}_\mathrm{BB}\end{array}\right),
\end{equation}
which yield the H and unpolarized fields as required, respectively; see Supplementary Material Figure~S2c,d.

\textit{Encoding scheme-3: Coherence-rank communications.} In encoding scheme-3, the logical bits are encoded in pairs, thus forming a quaternary code (an alphabet formed of 4 elements). Bit strings of 00, 01, 10, and 11 are encoded as R1 (rank-1), R2 (rank-2), R3 (rank-3), and R4 (rank-4) partially coherent fields, respectively. For each rank, we select the field configuration having the maximum entropy $S$ commensurate with the rank. In all cases, we prepare the fields in the diagonal representation. 
\begin{enumerate}
\item \textit{Rank-1.} The coherence matrix for the R1 field is:
\begin{equation}
\mathbf{G}_{00}=\mathrm{diag}\{1,0,0,0\}=\left(\begin{array}{cc}1&0\\0&0\end{array}\right)\otimes\left(\begin{array}{cc}1&0\\0&0\end{array}\right),    
\end{equation}
which is linearly polarized along H and is spatially coherent (mode $a$). The field is separable with respect to the spatial and polarization DoFs, and can be prepared from the source $\mathbf{G}_{\mathrm{o}}$ via the non-unitary transformation (Supplementary Material Figure~S3a):
\begin{equation}
\hat{T}_{1}=\left(\begin{array}{cc}\hat{T}_\mathrm{H}&\mathbf{0}\\\mathbf{0}&\hat{T}_\mathrm{BB}\end{array}\right).
\end{equation}

\item \textit{Rank-2.} The coherence matrix for the R2 field is:
\begin{equation}
\mathbf{G}_{01}=\frac{1}{2}\mathrm{diag}\{1,0,1,0\}=\left(\begin{array}{cc}1&0\\0&0\end{array}\right)\otimes\frac{1}{2}\left(\begin{array}{cc}1&0\\0&1\end{array}\right),
\end{equation}
which is linearly polarized along H but is spatially incoherent. The field is separable with respect to the spatial and polarization DoFs, and can be prepared from the source $\mathbf{G}_{\mathrm{o}}$ via the non-unitary transformation (Supplementary Material Figure~S3b):
\begin{equation}
\hat{T}_{2}=\left(\begin{array}{cc}\hat{T}_\mathrm{H}&\mathbf{0}\\\mathbf{0}&\hat{T}_\mathrm{H}\end{array}\right).    
\end{equation}

\item \textit{Rank-3.} The coherence matrix for the R3 field is:
\begin{equation}
\mathbf{G}_{10}=\frac{1}{3}\mathrm{diag}\{1,1,1,0\},
\end{equation}
which is partially polarized and spatially incoherent, but non-separable with respect to the spatial and polarization DoFs. The field can be prepared from the source $\mathbf{G}_{\mathrm{o}}$ via the non-unitary transformation (Supplementary Material Figure~S3c):
\begin{equation}
\hat{T}_{3}=\left(\begin{array}{cc}\mathbb{I}&\mathbf{0}\\\mathbf{0}&\hat{T}_\mathrm{H}\end{array}\right).
\end{equation}

\item \textit{Rank-4.} The coherence matrix for the R4 field is:
\begin{equation}
\mathbf{G}_{11}=\frac{1}{4}\mathrm{diag}\{1,1,1,1\}=\frac{1}{2}\left(\begin{array}{cc}1&0\\0&1\end{array}\right)\otimes\frac{1}{2}\left(\begin{array}{cc}1&0\\0&1\end{array}\right)=\mathbf{G}_{\mathrm{o}},
\end{equation}
which is unpolarized and spatially incoherent. The field is separable with respect to the spatial and polarization DoFs, and can be prepared from the source $\mathbf{G}_{\mathrm{o}}$ directly (Supplementary Material Figure~S3d):
\begin{equation}
\hat{T}_{4}=\left(\begin{array}{cc}\mathbb{I}&\mathbf{0}\\\mathbf{0}&\mathbb{I}\end{array}\right).
\end{equation}

\end{enumerate}

\noindent\textbf{Channel-1.} Channel-1 (Ch-1) shown in Fig.~\ref{Fig:MeasurementsCh1}(a) consists solely of a HWP that rotates from bit-to-bit at an angle located in a uniform random probability distribution consisting of integers between $0$ and $45^\circ$, see Supplementary Material Figure~S4a,b. In either case, the correlation lengths of $\theta$ are equal to $1$, see Supplementary Material Figure~S4d,e. To visualize the uniform random probability distribution from which the HWP samples, we plot the probability distribution of the random number generator when 10,000 numbers are generated in Supplementary Figure~S4c. Because the source is prepared as one spatial mode, the channel is represented by a $2\times2$ matrix $\hat{U}_{\mathrm{HWP}}(\theta)$ (Supplementary Material Section~1).

Simulations of Ch-1 shown in Supplementary Material Figure~S5 yield in $\mathrm{BER}=53\%$ for encoding scheme-1 and $\mathrm{BER}=0\%$ for encoding scheme-2. The cross-talk matrices used throughout this work are normalized such that the sum of each row is equal to 1. In the simulations plotted in Supplementary Material Figure~S5, the autocorrelation of $\theta$ drops below $0.5$ after the first bit, corresponding to a correlation lengths of $1$. The experimental results in Fig.~\ref{Fig:MeasurementsCh1} are in good agreement with the simulation in Supplementary Material Figure~S5.

\noindent\textbf{Channel-2.} The optical setup for Ch-2 sketched in Fig.~\ref{Fig:MeasurementsCh2}(a) consists of two fixed HWPs placed in the paths of the spatial modes $a$ and $b$, whose axes are rotated by $\tfrac{\pi}{8}$ with respect to H. Next, the two spatial modes $a$ and $b$ are superposed at a PBS, and the two output modes $a'$ and $b'$ pass through two HWPs that independently rotate between every bit to angles from a uniform random probability distribution consisting of integers between $0$ and $45^\circ$ (with respect to H), see Supplementary Material Figure~S6. All the equations for the field transformations associated with Ch-2 and the resulting coherence matrices are provided in Supplementary Material Section~4.

\noindent\textbf{Thresholding for bits.} In the three encoding schemes utilized, we make use of different thresholds to determine the bit values of the signal at the channel output.
\begin{enumerate}
    \item \textit{Threshold method for encoding in orthogonal polarization projections (encoding scheme-1).}  
    The polarization Stokes parameters were measured at the output of the channel to determine the state of polarization. The Stokes parameters are measured as follows: $S_{0}=I_\mathrm{H}+I_\mathrm{V}$ or the total intensity of the field; $S_{1}=I_\mathrm{H}-I_\mathrm{V}$ or the difference in intensity between the H and V projections; $S_{2}=2*I_\mathrm{D}-S_{0}$ or the difference between $+45^\circ$ and $-45^\circ$ polarized light; and $S_{3}=S_{0}-2*I_\mathrm{R}$ or the difference between right-handed circular polarization and left-handed circular polarization. The Stokes parameters were normalized by $S_{0}$.
    Because $S_{1}$ is the difference in intensity between H and V, the following scheme is used to discriminate between bits: if $S_{1}\geq0$, then the symbol is read as H (or bit of 0), if $S_{1}<0$, then the symbol is read as V (or bit of 1).
    
    \item \textit{Threshold method for encoding in the degree of polarization (encoding scheme-2).}
    Here the Stokes parameters are measured as described above, then the degree of polarization \textit{P} is calculated as $P=\sqrt{\frac{S_{1}^{2}+S_{2}^{2}+S_{3}^{2}}{S_0}}$. To discriminate between bits here the following scheme is used: if $P\geq0.5$ then the symbol is read as $P=1$ (or bit 0), if $P<0.5$ then the symbol is read as $P=0$ (or bit 1).  
    
    \item \textit{Threshold method for coherence-rank encoding (encoding scheme-3).} The $\mathbf{G}$ matrix is reconstructed for the field at the channel output using OCmT (Supplementary Material Section~5). To determine the eigenvalues of $\mathbf{G}$, a $4\!\times\!4$ unitary $\hat{U}$ spanning both DoFs \cite{Abouraddy17OE} diagonalizes $\mathbf{G}$: $\mathbf{G}_{\mathrm{D}}\!=\!\hat{U}\mathbf{G}\hat{U}^{\dag}\!=\!\mathrm{diag}\{\lambda_{1},\lambda_{2},\lambda_{3},\lambda_{4}\}$, with $\sum_{j}\lambda_{j}\!=\!1$. These eigenvalues are put in descending order by magnitude and then compared to the following `ideal' eigenvalues: for rank-1 $\mathbf{G}_\mathrm{D}=\mathrm{diag}\{1,0,0,0\}$, for rank-2 $\mathbf{G}_\mathrm{D}=\frac{1}{2}\mathrm{diag}\{1,1,0,0\}$, for rank-3 $\mathbf{G}_\mathrm{D}=\frac{1}{3}\mathrm{diag}\{1,1,1,0\}$, and rank-4 $\mathbf{G}_\mathrm{D}=\frac{1}{4}\mathrm{diag}\{1,1,1,1\}$, which correspond to the vertices of the 3D volume plotted in the restricted space spanned by $\{\lambda_{1},\lambda_{2},\lambda_{3}\}$ (Fig.~\ref{Fig:CoherenceRank}. To discriminate between coherence ranks for fields output from the channels, the measured eigenvalues are plotted in the restricted $\{\lambda_{1},\lambda_{2},\lambda_{3}\}$-space in Fig.~\ref{Fig:CoherenceRank}(d), and their coordinates are compared to each vertex of the 3D structure via the Euclidean metric. 
    
    The Euclidean distance from the rank-1 vertex is $\sqrt{(\lambda_{1}-1)^{2}+\lambda_{2}^{2}+\lambda_{3}^{2}+\lambda_{4}^{2}}$, the Euclidean distance from the rank-2 vertex is $\sqrt{(\lambda_1-\tfrac{1}{2})^{2}+(\lambda_{2}-\tfrac{1}{2})^{2}+\lambda_{3}^{2}+\lambda_{4}^{2}}$, the Euclidean distance from the rank-3 vertex is $\sqrt{(\lambda_{1}-\tfrac{1}{3})^{2}+(\lambda_{2}-\tfrac{1}{3})^{2}+(\lambda_{3}-\tfrac{1}{3})^{2}+\lambda_{4}^{2}}$, and the Euclidean distance from the rank-4 vertex is $\sqrt{(\lambda_{1}-\tfrac{1}{4})^{2}+(\lambda_{2}-\tfrac{1}{4})^{2}+(\lambda_{3}-\tfrac{1}{4})^{2}+(\lambda_{4}-\tfrac{1}{4})^{2}}$. The coherence rank associated with the vertex that yields the minimum Euclidean distance from the measured eigenvalues is read as the output symbol.
\end{enumerate}


\section{Matrix representation of the optical devices used in the communications channels}

We present here the matrix representation of the optical devices and systems used in the synthesis of different-rank optical fields and in the communications channels described in the main text.

\subsection{Definition of the coherence matrix}

The $4\times4$ coherence matrix $\mathbf{G}$ (which is Hermitian, unity-trace, and positive semi-definite) encompassing the polarization and spatial degrees-of-freedom (DoFs) is defined as follows: 
\begin{equation}
\mathbf{G} =
\begin{pmatrix}
G^\mathrm{HH}_{aa} & G^\mathrm{HV}_{aa} & G^\mathrm{HH}_{ab} & G^\mathrm{HV}_{ab}\\
G^\mathrm{VH}_{aa} & G^\mathrm{VV}_{aa} & G^\mathrm{VH}_{ab} & G^\mathrm{VV}_{ab}\\
G^\mathrm{HH}_{ba} & G^\mathrm{HV}_{ba} & G^\mathrm{HH}_{bb} & G^\mathrm{HV}_{bb}\\
G^\mathrm{VH}_{ba} & G^\mathrm{VV}_{ba} & G^\mathrm{VH}_{bb} & G^\mathrm{VV}_{bb}\\
\end{pmatrix},
\label{eq:G}
\end{equation}
where $G^{jl}_{km}=\langle{E^j_k(E^l_m)^*}\rangle$, $j,l=$H,V, and $k,m=a,b$ \cite{Abouraddy14OL,Kagalwala15SR,Harling23JO}. The polarization DoF is spanned by the orthogonal H (horizontal) and V (vertical) polarization components, and the spatial DoF is spanned by the two spatial modes identified by $a$ and $b$.

Another coherence matrix of interest is the reduced polarization coherence matrix $\mathbf{G}_{\mathrm{p}}^{\mathrm{red.}}$ obtained from $\mathbf{G}$ by tracing over the spatial DoF:
\begin{equation}
\mathbf{G}_{\mathrm{p}}^{\mathrm{red.}}=\left(\begin{array}{cc}G^\mathrm{HH}_{bb}+G^\mathrm{HH}_{aa}&G^\mathrm{HV}_{aa}+G^\mathrm{HV}_{bb}\\G^\mathrm{VH}_{aa}+G^\mathrm{VH}_{bb}&G^\mathrm{VV}_{aa}+G^\mathrm{VV}_{bb}\end{array}\right).
\end{equation}

In general, a deterministic optical system operating on this field is represented by a $4\times4$ transformation matrix $\hat{T}$ (whether unitary or non-unitary), such that the resulting coherence matrix $\mathbf{G}`$ is given by: $\mathbf{G}`\!=\!\hat{T}\mathbf{G}\hat{T}^{\dag}$ (Fig.~\ref{fig:GeneralTransformations}a). 

\subsection{Polarization devices}

Any polarization transformation operating on one spatial mode is represented by a $2\times2$ Jones matrix. 

\begin{enumerate}
\item \textit{Half-wave plate.} For a half-wave plate (HWP), the unitary matrix representation is:
\begin{equation}\hat{U}_{\mathrm{HWP}}(\theta)\!=\!\left(\begin{array}{cc}
\cos2\theta&\sin2\theta\\\sin2\theta&-\cos2\theta\end{array}\!\right),\end{equation} where $\theta$ is the angle of the fast axis of the HWP with respect to H. 
\item \textit{Linear polarizer.} The non-unitary operator for projecting polarization onto a linear basis rotated by an angle $\psi$ with respect to H:
\begin{equation}
\hat{T}_{\mathrm{P}}(\psi)=\left(\begin{array}{cc}\cos^{2}\psi&\cos\psi\sin\psi\\\cos\psi\sin\psi&\sin^{2}\psi\end{array}\right).
\end{equation}
For linear polarizers along H and V we have:
\begin{equation}
\hat{T}_{\mathrm{H}}=\hat{T}_{\mathrm{P}}(0)=\left(\begin{array}{cc}1&0\\0&0\end{array}\right),\;\;\;\;
\hat{T}_{\mathrm{V}}=\hat{T}_{\mathrm{P}}(90^{\circ})=\left(\begin{array}{cc}0&0\\0&1\end{array}\right), 
\end{equation}
respectively.
\item \textit{Beam block (BB).} When one path of a spatial mode is blocked, we have the transformation: 
\begin{equation}
\hat{T}_{\mathrm{BB}}=\left(\begin{array}{cc}
0&0\\0&0\end{array}\!\right)=\mathbf{0}
\end{equation}

\end{enumerate}

\subsection{Separable transformations}

\begin{figure*}[t!]
    \centering
    \includegraphics[width=12cm]{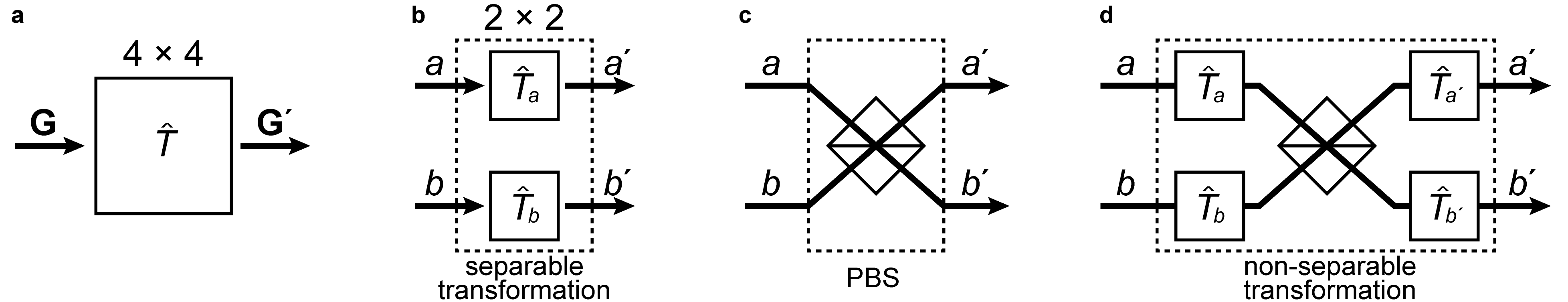}
    \caption{\textbf{Transformations of the optical field across both DoFs.} (a) A general transformation $\hat{T}$ (represented by a $4\times4$ matrix) transforming the coherence matrix $\mathbf{G}$ to $\mathbf{G}'$. (b) A separable transformation implements two polarization transformations $\hat{T}_{a}$ and $\hat{T}_{b}$ (each represented by a $2\times2$ matrix) to the spatial modes $a$ and $b$, respectively; see Eq.~\ref{eq:Separable}. (c) A polarizing beam splitter (PBS) coupling the spatial and polarization DoFs; see Eq.~\ref{eq:PBS}. (d) A general non-separable transformation formed of a separable transformation, followed by a PBS and then a second separable transformation; see Eq.~\ref{eq:Nonseparable}.}
    \label{fig:GeneralTransformations}
\end{figure*}

Separable transformations take the form of two polarization devices $\hat{T}_{a}$ and $\hat{T}_{b}$, each operating separately on one of the two spatial modes $a$ and $b$, respectively, as shown in Fig.~\ref{fig:GeneralTransformations}b. The overall $4\times4$ transformation $\hat{T}$ is then given by the block-diagonal form:
\begin{equation}\label{eq:Separable}
\hat{T}\!=\!\left(\begin{array}{cc}\hat{T}_{a}&\mathbf{0}\\\mathbf{0}&\hat{T}_{b}\end{array}\right), 
\end{equation}
where $\mathbf{0}$ is a $2\times2$ matrix of zeros. Non-separable transformations cannot be written in block-diagonal form.

\subsection{Non-separable transformations}

Non-separable transformations in general couple the spatial and polarization DoFs, and thus cannot be written in block-diagonal form. An example is a polarizing beam splitter (PBS) that introduces a polarization-sensitive superposition of the spatial modes, and is thus non-separable. The PBS used in our work transmits H and reflects V from either $a$ or $b$. The corresponding unitary matrix representation is thus given by (Fig.~\ref{fig:GeneralTransformations}c):
\begin{equation}\label{eq:PBS}
\hat{U}_{\mathrm{PBS}}=\left(\begin{array}{cccc}
0&0&1&0\\
0&i&0&0\\
1&0&0&0\\
0&0&0&i\end{array}\!\right).
\end{equation}
More generally, the configuration shown in Fig.~\ref{fig:GeneralTransformations}d that comprises separable, polarization transformations $\hat{T}_{a}$ and $\hat{T}_{b}$, followed by a PBS and then two other separable polarization transformations $\hat{T}_{a}'$ and $\hat{T}_{b}'$ is given overall by the $4\times4$ unitary matrix transformation:
\begin{equation}\label{eq:Nonseparable}
\hat{U}=\left(\begin{array}{cc}\hat{T}_{a}'&\mathbf{0}\\\mathbf{0}&\hat{T}_{b}'\end{array}\right)\hat{U}_{\mathrm{PBS}}\left(\begin{array}{cc}\hat{T}_{a}&\mathbf{0}\\\mathbf{0}&\hat{T}_{b}\end{array}\right).
\end{equation}

\section{Source preparation and encoding schemes}

\begin{figure*}[h!]
    \centering
    \includegraphics[width=12cm]{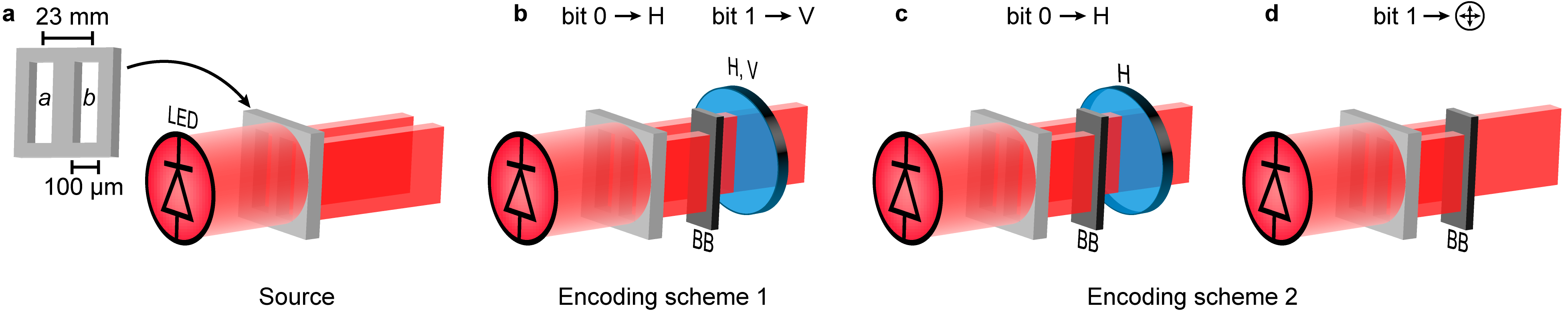}
    \caption{\textbf{Encoding schemes for the optical source.} (a) Schematic of the optical source, comprising an LED from which two spatial modes $a$ and $b$ are extracted via a pair of slits (inset). (b) Schematic of the optical configurations for encoding scheme~1. (c,d) Schematic of the optical configurations for encoding scheme-2. BB: Beam block; H: linear polarizer along H; V: linear polarizer along V.}
    \label{fig:EncodingForCh1}
\end{figure*}

\begin{figure*}[h!]
    \centering
    \includegraphics[width=12cm]{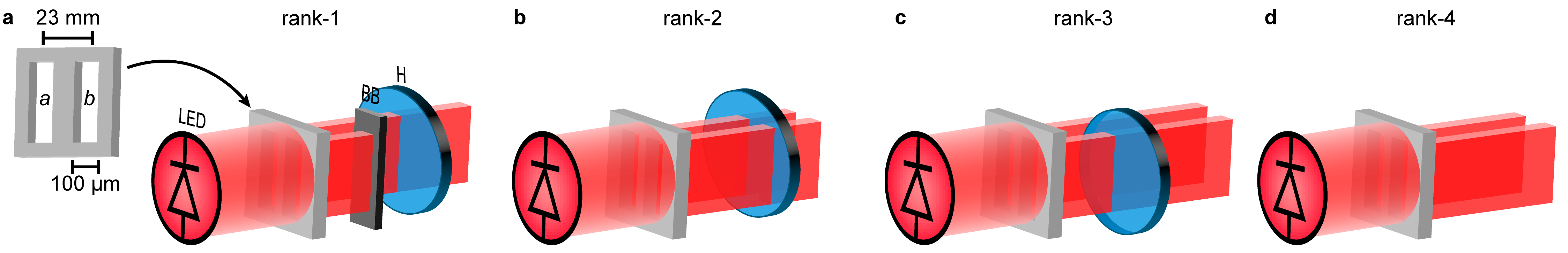}
    \caption{\textbf{Source configurations for coherence-rank encoding.} The four coherence ranks, rank-1 (R1) through rank-4 (R4), are prepared starting from a spatially incoherent, unpolarized LED. (a) Synthesizing rank-1 where \textit{a} is horizontally polarized and $b$ is blocked, (b) rank-2 where $a$ and $b$ are horizontally polarized, (c) rank-3 where $b$ is horizontally polarized while $a$ is unpolarized, and (d) rank-4 where $a$ and $b$ are both unpolarized. H: Linear polarizer along the horizontal; BB: beam block. The inset in (a) shows the dimensions of the double-slits used to produce the two spatial modes $a$ and $b$.}
    \label{fig:Rank sources}
\end{figure*}

Schematic depictions of the optical source preparation, in addition to the experimental configurations for encoding scheme-1 and encoding scheme-2 are provided in Fig.~\ref{fig:EncodingForCh1}. See Methods in main text for details. Coherence-rank communications makes use of encoding scheme-3, and the schematics for the configurations that produce the different-ranked fields are depicted in Fig.~\ref{fig:Rank sources}. See Methods in main text for details.

\section{Channel-1}

\begin{figure*}[t!]
    \centering
    \includegraphics[width=12cm]{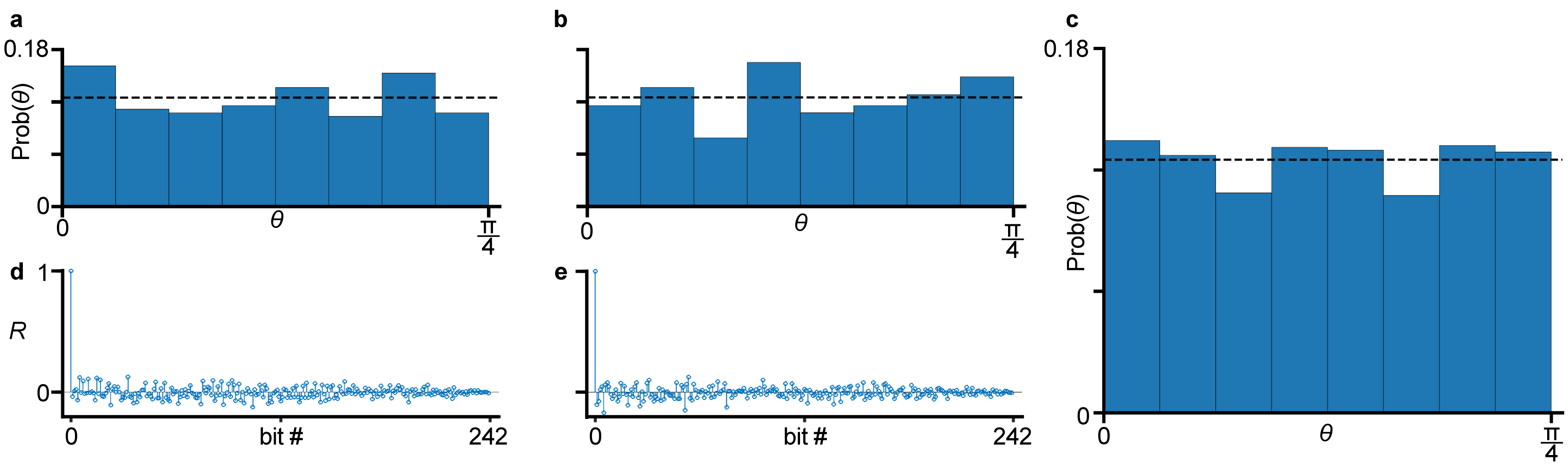}
    \caption{\textbf{The probability distributions and correlation functions for Ch-1.} (a) The experimental probability distribution $\mathrm{Prob}(\theta)$ for the randomly selected angle $\theta$ of the HWP in Ch-1 when employing encoding scheme-1. Here 242~angles are generated in the range $[0,\frac{\pi}{4}]$, collected in 8~bins, and normalized such that $\sum_{j=1}^{8}\mathrm{Prob}(\theta_{j})\!=\!1$, where $j$ indicates the angular bin. (b) Same as (a) for encoding scheme-2. (c) Theoretical version of (a,b) for 10,000 angles generated. The horizontal dashed lines in (a--c) are the expected averages. (d) The autocorrelation function for $\theta$, $R(n)\!=\!\sum_{j=1}^{242}(\theta_{j}-\tfrac{\pi}{8})(\theta_{j+n}-\tfrac{\pi}{8})$ for encoding scheme-1 and (e) encoding scheme-2.}
    \label{fig:Ch-1 distribution}
\end{figure*}

\begin{figure*}[t!]
    \centering
    \includegraphics[width=17cm]{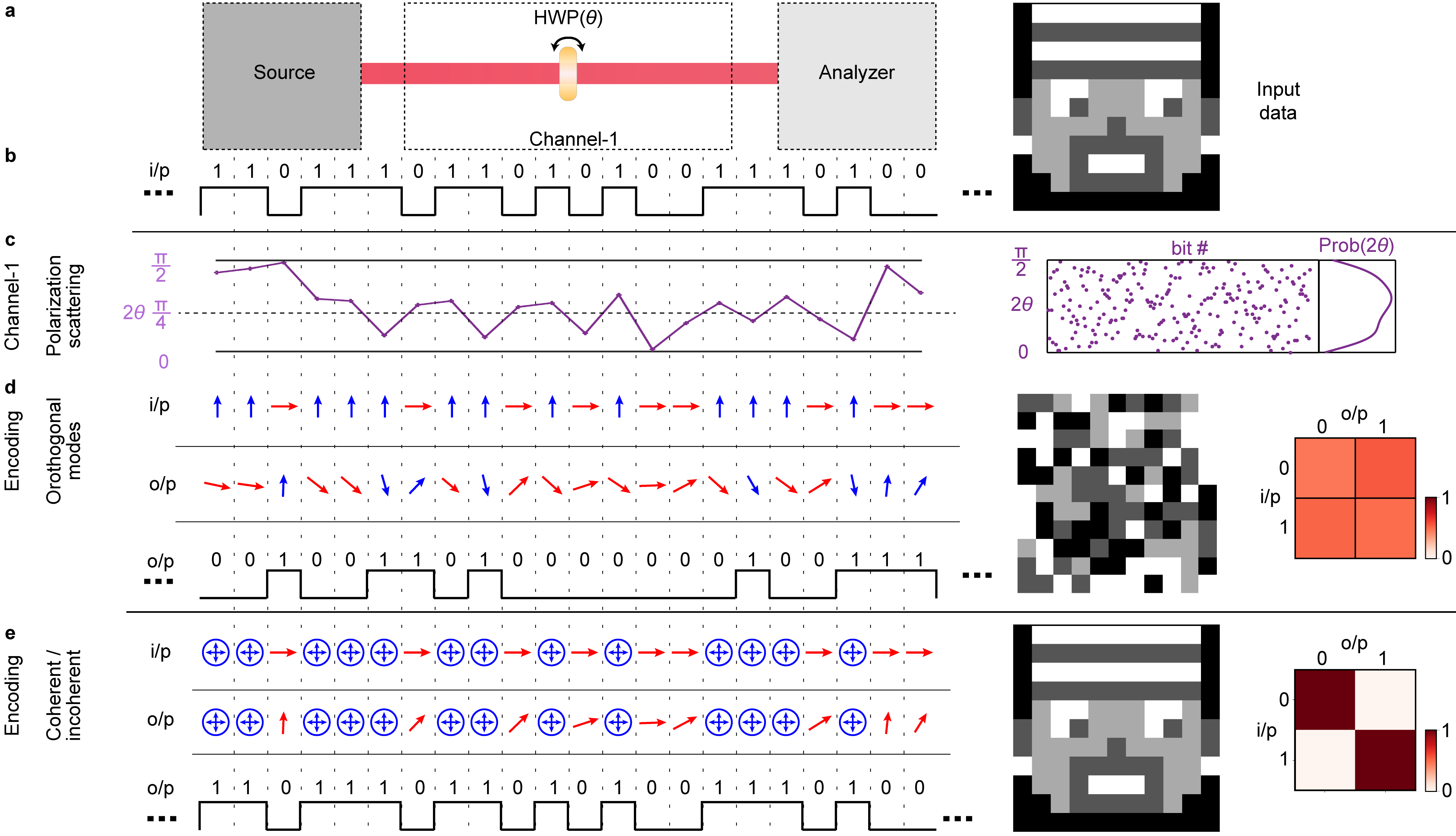}
    \caption{\textbf{Simulations for encoding scheme-1 and encoding scheme-2 through Ch-1.} (a) Schematic of the setup for Ch-1. (b) Portion of the data stream corresponding to the input image on the right. (c,d) Polarization encoding scheme-1, where the logical bits are mapped to orthogonal linear polarization states: $0\rightarrow$H and $1\rightarrow$V. (c) Settings for the HWP angle $\theta$ in the channel. The probability distribution of $\theta$ is plotted on the right. (d) Input and output polarization states corresponding to the data stream. The cross-talk matrix is plotted on the right, along with the reconstructed image. (e) Same as (d) for polarization encoding scheme-2, where the logical bits are mapped to polarized and unpolarized fields: $0\rightarrow$H and $1\rightarrow$ \protect\unpolarizedSymbol, repectively.}
    \label{fig:channel 1 simulation}
\end{figure*}

We plot in Fig.~\ref{fig:Ch-1 distribution} the measured probability distributions and correlation functions for the angular settings of the HWP in Ch-1 (Fig.~5a in the main text). Simulations of Ch-1 are presented in Fig.~\ref{fig:channel 1 simulation}. See Methods for details.

\section{Channel-2}

The general matrix representation for Ch-2 is:
\begin{eqnarray}
\mathbf{T}&=&
\resizebox{.35\hsize}{!}{
$\left(\begin{array}{cc}\hat{U}_{\mathrm{HWP}}(\varphi)&\mathbf{0}\\\mathbf{0}&\hat{U}_{\mathrm{HWP}}(\theta)\end{array}\right)$}
\hat{U}_{\mathrm{PBS}}
\resizebox{.35\hsize}{!}{
$\left(\begin{array}{cc}\hat{U}_{\mathrm{HWP}}(\pi/8)&\mathbf{0}\\\mathbf{0}&\hat{U}_{\mathrm{HWP}}(\pi/8)\end{array}\right)$}\nonumber\\
&=&\frac{1}{\sqrt{2}}
\resizebox{.70\hsize}{!}{
$\left(\begin{array}{cccc}
i\sin(2\varphi)&-i\sin(2\varphi)&\cos(2\varphi)&\cos(2\varphi)\\
-i\cos(2\varphi)&i\cos(2\varphi)&\sin(2\varphi)&\sin(2\varphi)\\
\cos(2\theta)&\cos(2\theta)&i\sin(2\theta)&-i\sin(2\theta)\\
\sin(2\theta)&\sin(2\theta)&-i\cos(2\theta)&i\cos(2\theta)
\end{array}\right)$}.
\end{eqnarray}

\begin{figure*}[t!]
    \centering
    \includegraphics[width=12cm]{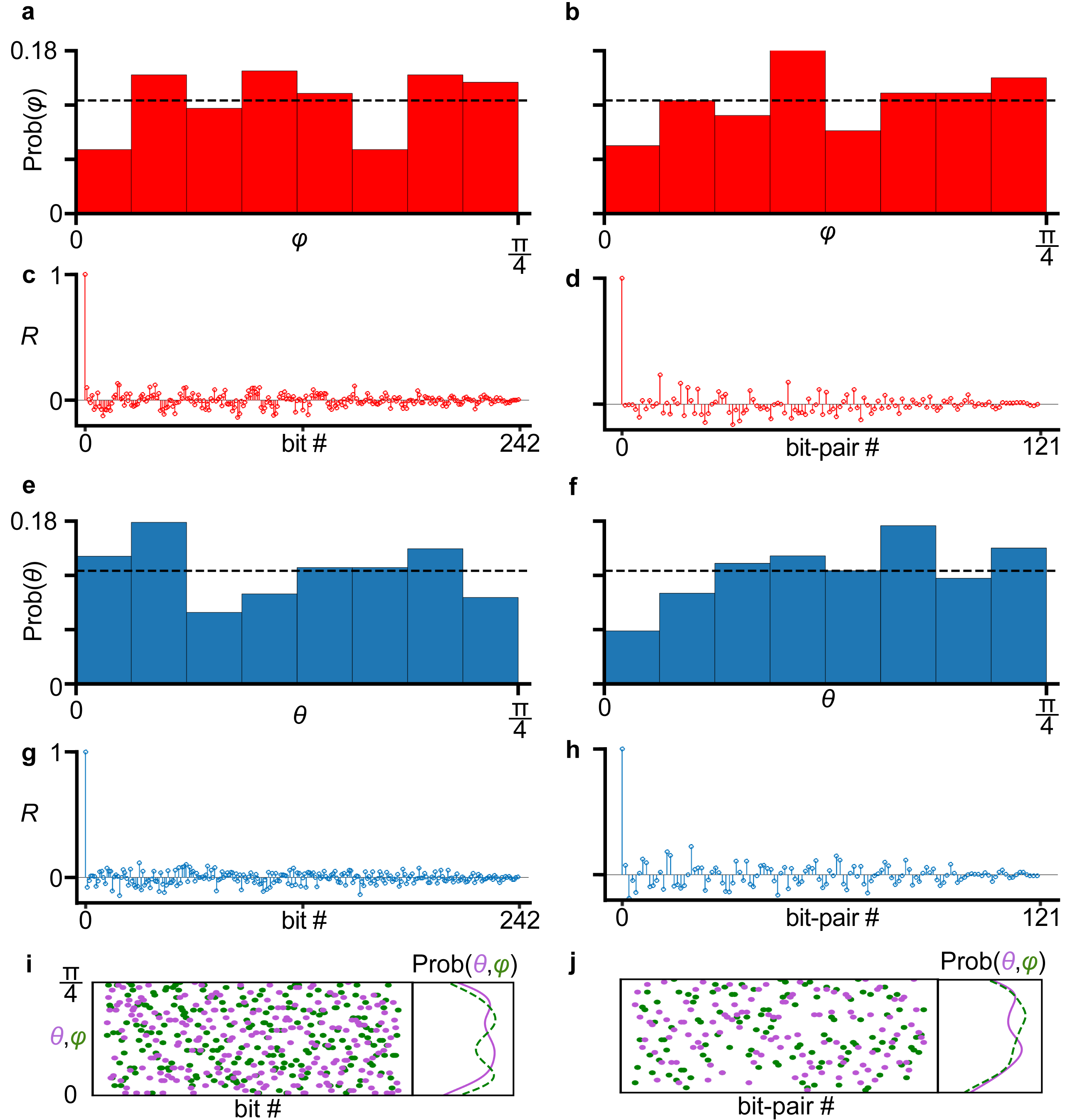}
    \caption{\textbf{Probability distributions and correlations functions for Ch-2.} (a) The experimental probability distribution $\mathrm{Prob}(\varphi)$ for the randomly selected angle $\varphi$ of the HWP in Ch-2 when employing encoding scheme~2, and (b) encoding scheme-3. Here 242 and 121 angles for encoding scheme-2 and encoding scheme-3, respectively, are generated in the range $[0,\frac{\pi}{4}]$, collected in 8~bins, and normalized such that $\sum_{j=1}^{8}\mathrm{Prob}(\varphi_{j})\!=\!1$, where $j$ indicates the angular bin. (c) The autocorrelation function of $\varphi$, $R(n)\!=\!\sum_{j=1}^{242}(\varphi_{j}-\tfrac{\pi}{8})(\varphi_{j+n}-\tfrac{\pi}{8})$ for encoding scheme-2 and (d) encoding scheme-3. (e--h) Same as (a--d) for $\theta$. The horizontal dashed lines in (a, b, e, f) are the expected averages.}
    \label{fig:Ch-2 distribution}
\end{figure*}

A transformation of $\mathbf{T}$ applied to $\mathbf{G}$ yields $\mathbf{G}'\!=\!\mathbf{T}\mathbf{G}\mathbf{T}^{\dag}$. From this the reduced polarization coherence matrix $\mathbf{G'}_\mathrm{p}$ is obtained by tracing out the spatial DoF. Here $\varphi$ and $\theta$ are the angles of the fast-axes of the half-wave plates with respect to the horizontal that are used in Ch-2. For encoding scheme-2 and encoding scheme-3 passing through Ch-2, see Fig.~\ref{fig:Ch-2 distribution}a,b,e,f for probability distributions of the half-wave plate angles used. The correlation lengths of $\varphi$ and $\theta$ are both equal to $1$~bit. For encoding-scheme-3, the sources were prepared as unpolarized $\mathbf{G}=\frac{1}{2}\mathrm{diag}\{1,1,0,0\}$ or polarized $\mathbf{G}=\mathrm{diag}\{1,0,0,0\}$:
\begin{equation}
\mathbf{G'}_\mathrm{p}=\frac{1}{4}
\resizebox{.87\hsize}{!}{
$\left(\begin{array}{cc}
2\left\{\cos^{2}(2\theta)+\sin^{2}(2\varphi)\right\} & \sin(4\theta)-\sin(4\varphi)\\
\sin(4\theta)-\sin(4\varphi)& 
2\left\{\cos^{2}(2\varphi)+\sin^{2}(2\theta)\right\}.
\end{array}\right)$}.
\label{eq: G_P after T}
\end{equation}   

Here the physical information that encodes bits is completely lost because $\mathbf{G'}_\mathrm{p}$ is the same whether horizontally polarized light or unpolarized light is sent into the channel.

For the sources prepared with the four ranks:
\begin{enumerate}
\item\textit{rank-1.} When $\mathbf{G}=\mathrm{diag}\{1, \;\; 0, \;\; 0, \;\; 0\}$, the full $4\times4$ coherence matrix is:  
\begin{equation}
\mathbf{G'}\!=\!\frac{1}{2}
\resizebox{.87\hsize}{!}{
$\left(\begin{array}{cccc}
\sin^{2}(2\varphi)&
-\tfrac{1}{2}\sin(4\varphi)& i\cos(2\theta)\sin(2\varphi)& i\sin(2\varphi)\sin(2\theta)\\
-\tfrac{1}{2}\sin(4\varphi)& \cos^{2}(2\varphi)&
-i\cos(2\varphi)\cos(2\theta)&
-i\cos(2\varphi)\sin(2\theta)\\
-i\cos(2\theta)\sin(2\varphi)& i\cos(2\varphi)\cos(2\theta)& \cos^{2}(2\theta)& \tfrac{1}{2}\sin(4\theta)\\
-i\sin(2\varphi)\sin(2\theta)&
i\cos(2\varphi)\sin(2\theta)&
\tfrac{1}{2}\sin(4\theta)&
\sin^{2}(2\theta) 
\end{array}\right)$}, 
\end{equation}
and the polarization coherence matrix reduces to Eq.~\ref{eq: G_P after T}.

\item\textit{rank-2.} When $\mathbf{G}=\frac{1}{2}\mathrm{diag}\{1, \;\; 0, \;\; 1, \;\; 0\}$, the full $4\times4$ coherence matrix is: 
\begin{equation}
\mathbf{G'}\!=\!\frac{1}{4}
\resizebox{.87\hsize}{!}{
$\left(\begin{array}{cccc}
1 & 0 & i\sin(2\varphi-2\theta)&
i\cos(2\varphi-2\theta)\\
0 & 1 & -i\cos(2\varphi-2\theta)&
i\sin(2\varphi-2\theta)\\
-i\sin(2\varphi-2\theta)&i\cos(2\varphi-2\theta)&1&0\\
-i\cos(2\varphi-2\theta)&-i\sin(2\varphi-2\theta)&0&1
\end{array}\right)$}
\end{equation}
and the polarization coherence matrix reduces to $\mathbf{G'}_\mathrm{p}=\tfrac{1}{2}
\left(\begin{array}{cc}1&0\\0&1\end{array}\right)$.

\begin{figure*}[t!]
    \centering
    \includegraphics[width=17cm]{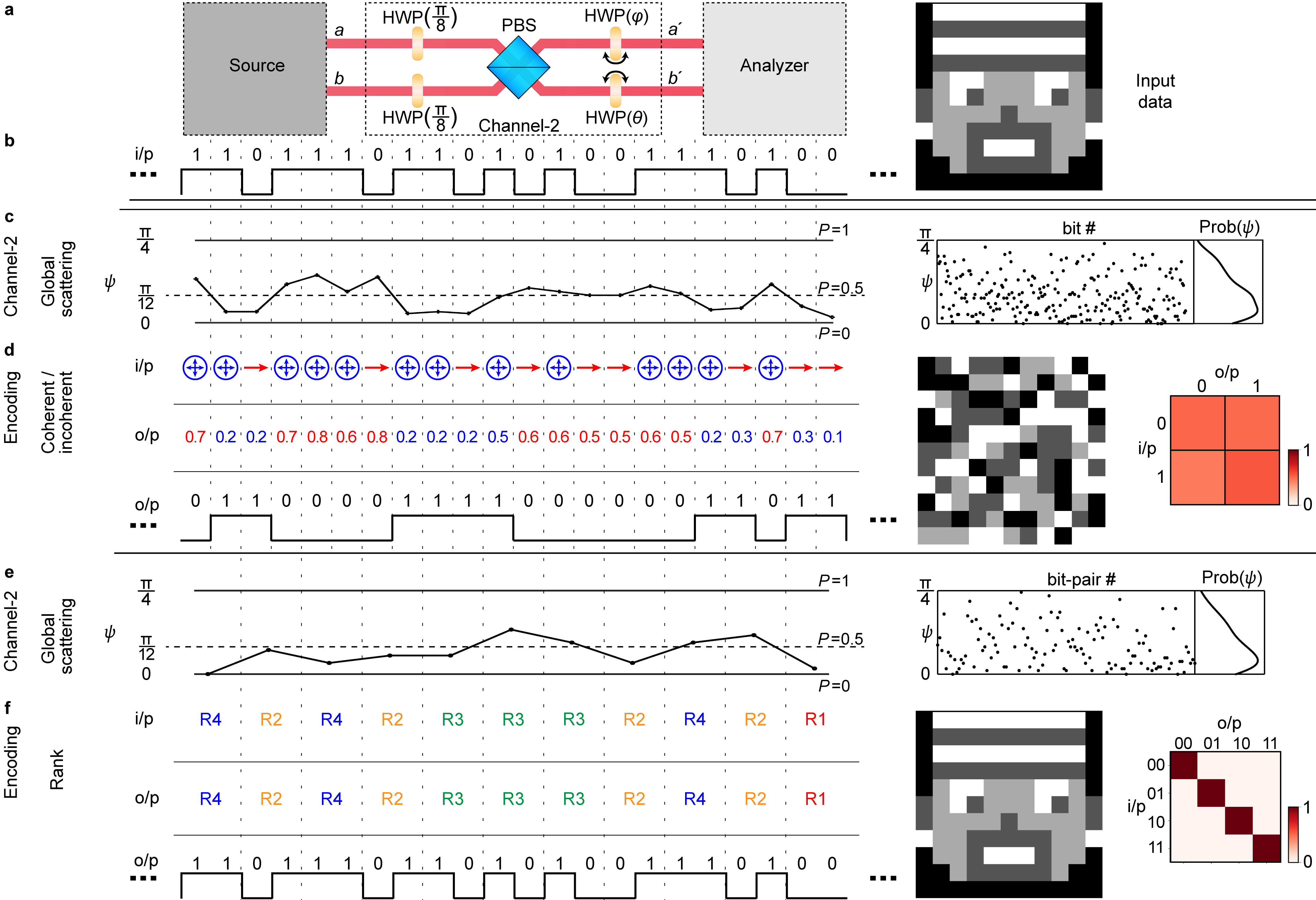}
    \caption{\textbf{Simulations for encoding scheme-2 and encoding scheme-3 through Ch-2.} (a) Schematic of the setup for Ch-2. (b) Portion of the data stream corresponding to the input image on the right. (c,d) Polarization encoding scheme-2: $0\rightarrow$H and $1\rightarrow$\protect\unpolarizedSymbol (unpolarized). (c) Settings for $\psi\!=\!|\theta-\varphi|$, where $\theta$ and $\varphi$ are the HWP angles in the channel, and the probability distribution of $\psi$ is plotted on the right. (d) Input and output polarization states corresponding to the data stream. The measured cross-talk matrix is plotted on the right, along with the reconstructed image. (e,f) Same as (c,d) for encoding scheme-3, corresponding to coherence-rank communications. Pairs of logical bits are encoded in the coherence rank: $00\rightarrow$R1, $01\rightarrow$R2, $10\rightarrow$R3, and $11\rightarrow$R4. (f) The expected input and output coherence ranks.}
    \label{fig:channel 2 simulation}
\end{figure*}

\item\textit{rank-3.} When $\mathbf{G}=\frac{1}{3}\mathrm{diag}\{1, \;\; 1, \;\; 1, \;\; 0\}$, the full $4\times4$ coherence matrix is
\begin{equation}
\mathbf{G'}\!=\!\frac{1}{6} 
\resizebox{.87\hsize}{!}{$\left(\begin{array}{cccc}
\sin^{2}(2\varphi)+1&
-\tfrac{1}{2}\sin(4\varphi)&
-i\cos(2\varphi)\sin(2\theta)&
\cos(2\varphi)\cos(2\theta)\\
-\tfrac{1}{2}\sin(4\varphi)&
2-\sin^{2}(2\varphi)&
-i\sin(2\varphi)\sin(2\theta)&
i\cos(2\theta)\sin(2\varphi)\\
i\cos(2\varphi)\sin(2\theta)&
i\sin(2\varphi)\sin(2\theta)&
2-\sin^{2}(2\theta)&
\tfrac{1}{2}\sin(4\theta)\\
-\cos(2\varphi)\cos(2\theta)&
-i\cos(2\theta)\sin(2\varphi)&
\tfrac{1}{2}\sin(4\theta)&
\sin^{2}(2\theta)+1
\end{array}\right)$},
\end{equation}
and the polarization coherence matrix reduces to \begin{equation}
\mathbf{G'}_\mathrm{p}=\frac{1}{12}
\left(\begin{array}{cc}
\cos(4\theta)-\cos(4\varphi)+6&
\sin(4\theta)-\sin(4\varphi)\\
\sin(4\theta)-\sin(4\varphi)&
\cos(4\varphi)-\cos(4\theta)+6
\end{array}\right).
\end{equation}

\item\textit{rank-4.} When $\mathbf{G}=\frac{1}{4}\mathrm{diag}\{1, \;\; 1, \;\; 1, \;\; 1\}$, the polarization coherence matrix reduces to $\mathbf{G'}_\mathrm{p}=\frac{1}{2}
\left(\begin{array}{cc}1&0\\0&1\end{array}\right)$, and $\mathbf{G'}\!=\!\mathbf{G}$. In each case, the rank of $\mathbf{G}$ is equal to that of $\mathbf{G'}$. 

\end{enumerate}

The experimental results from Fig.~6 in the main text are in good agreement with the simulation in Fig. \ref{fig:channel 2 simulation}. The simulation of Ch-2 in Fig.~\ref{fig:channel 2 simulation} results in $\mathrm{BER} = 51 \%$ for encoding scheme-2 and $\mathrm{BER} = 0 \%$ for encoding scheme-3. For the simulation in Fig.~\ref{fig:channel 2 simulation}, the autocorrelations of $\varphi$ and $\theta$ drop below $0.5$ after the first bit,
resulting in correlation lengths for $\varphi$ and $\theta$ both equal to $1$. 

\begin{figure*}[t!]
    \centering
    \includegraphics[width=12cm]{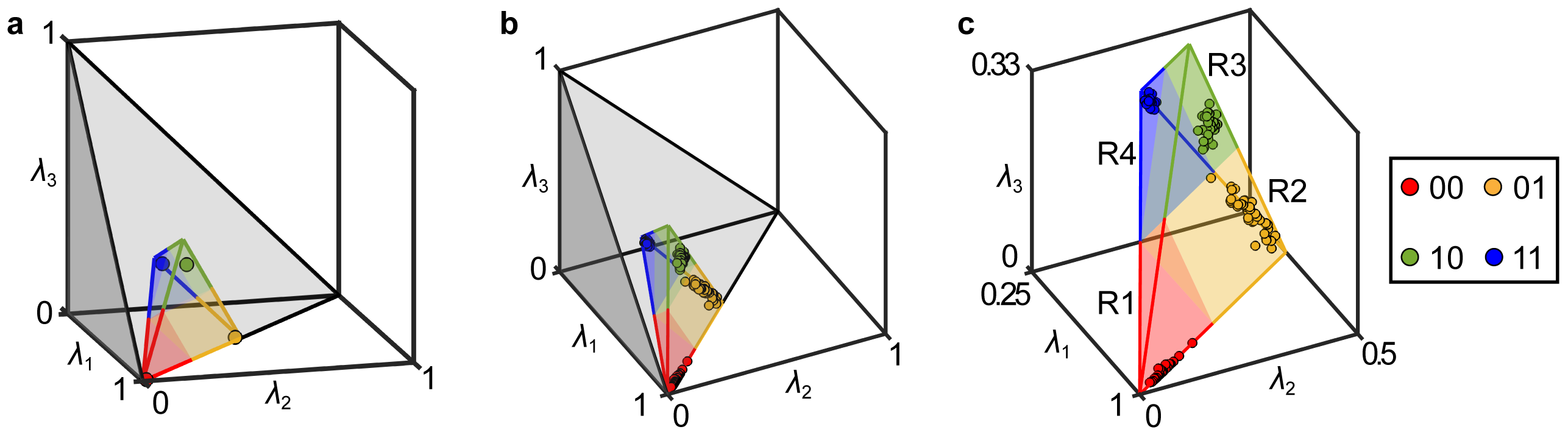}
    \caption{\textbf{Circumventing scattering by coherence-rank communications.} (a) The experimental sources for rank-1 to rank-4 are plotted in terms of their eigenvalues in the 3D eigenvalue space. (b) The output experimental eigenvalues from Fig.~6f in the main text in the full 3D eigenvalue space. (c) Restricting the geometry to the 3D eigenvalue subspace where eigenvalues are in descending order and zooming in to better visualize the experiment outcomes; the colored volumes correspond to the Euclidean distance thresholding.}
    \label{fig:rank subspace experiment distribution}
\end{figure*}

Figure~\ref{fig:rank subspace experiment distribution} shows the output of the rank in the eigenvalue 3D space for the encoding experiment. Experimental error led to spread in the eigenvalues. Due to the thresholding based on Euclidean distance, there is no bit error. 



\section{Tomographic reconstruction of coherence matrices}
\label{OCmT}

We make use of optical coherence matrix tomography (OCmT) \cite{Abouraddy02OptComm,Abouraddy14OL,Kagalwala15SR} to reconstruct the $4\times4$ coherence matrices $\mathbf{G}$.

\begin{enumerate}
 
\item\textit{Intensity measurements for OCmT reconstruction.} A linear polarizer provides projections along three directions: H, V, and D (diagonal, $45^{\circ}$). Adding a quarter wave plate, a projection in the circular polarization basis (R) is obtained. See Fig. \ref{fig:OCmT} for the full set of measurement configurations. We denote the intensity measurements $I^{j}_{k}$, where $j\!=$ H, V, D, and R represent the polarization projections, while $k\!=\!a,b,$ $a+b$, and $a+ib$ are the spatial projections. The intensity corresponding to the spatial projection $a$ is obtained by blocking the slit $b$ and measuring the peak intensity at the center of the diffraction pattern; and similarly for the spatial projection $b$. The spatial projection $a+b$ is obtained by allowing light to pass through both slits and measuring the intensity at the center of the interference pattern. The spatial projection $a+ib$ is the intensity at the location midway between the central peak and the first dip (minimum) of the interference pattern. In the case of an interference pattern with $0 \%$ visibility, $a+ib$ is the intensity at the expected location midway between the central peak and the first dip. A spherical lens (not shown in Fig. \ref{fig:OCmT}) of focal length $f\!=\!30$~mm is used to overlap the spatial modes from $a$ and $b$ at the Fourier plane, whereby interference fringes are observed if the field is not spatially incoherent. The full list of acquired intensity measurements is provided in Table~\ref{Table:OCmT}.

\begin{figure*}[t!]
    \centering
    \includegraphics[width=12cm]{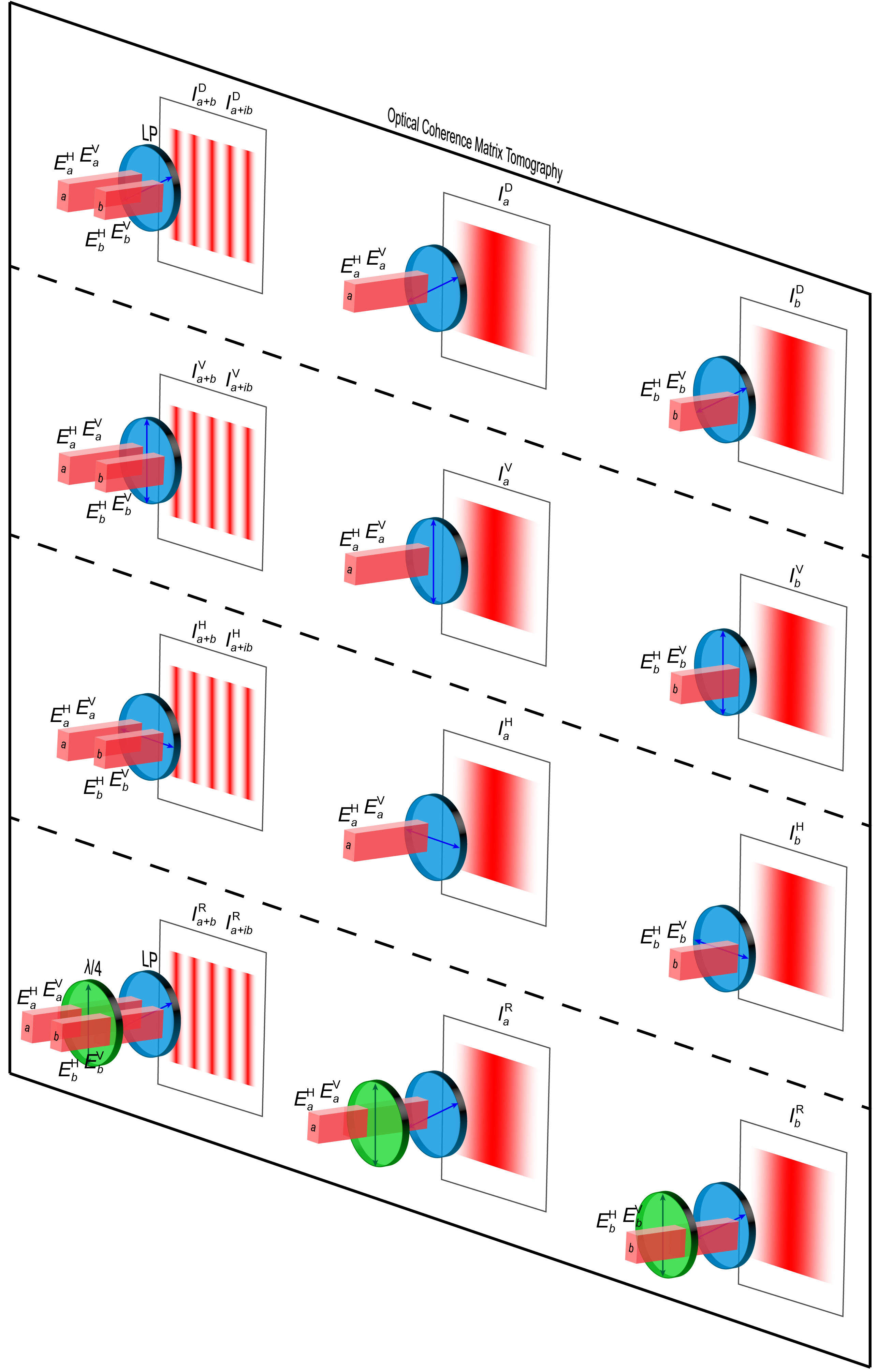}
    \caption{\textbf{Intensity projections for OCmT.} The measurements of the intensity projections are carried out sequentially for an incoming field to reconstruct the associated coherence matrices $\mathbf{G}$ via OCmT. LP: Linear polarizer; $\lambda/4$: quarter wave plate.}
    \label{fig:OCmT}
\end{figure*}

\begin{table}[b!]
    \centering
    \caption{The tomographic intensity measurements necessary for  OCmT. H: Horizontal; V: vertical; D: $45^{\circ}$; R: right-hand circular.}
    \begin{tabular}{|l|*{6}{c|}}
    \hline \multicolumn{1}{|c}{}& &\multicolumn{4}{c|}{\textbf{polarization projection}}\\
    \cline{3-6} \multicolumn{1}{|c}{}&\multicolumn{1}{c|}{} &H & V & D & R\\ 
    \hline \multirow{4}{*}{\centering{\textbf{spatial projection}}} &$a$ & $I^{\mathrm{H}}_{a}$ & $I^{\mathrm{V}}_{a}$ & $I^{\mathrm{D}}_{a}$ &$I^{\mathrm{R}}_{a}$ \\
    \cline{2-6} &$b$ & $I^{\mathrm{H}}_{b}$ & $I^{\mathrm{V}}_{b}$ & $I^{\mathrm{D}}_{b}$ & $I^{\mathrm{R}}_{b}$\\
    \cline{2-6} &$a+b$ & $I^{\mathrm{H}}_{(a+b)}$ & $I^{\mathrm{V}}_{(a+b)}$ & $I^{\mathrm{D}}_{(a+b)}$ & $I^{\mathrm{R}}_{(a+b)}$\\
    \cline{2-6} &$a+ib$ & $I^{\mathrm{H}}_{(a+ib)}$ & $I^{\mathrm{V}}_{(a+ib)}$ & $I^{\mathrm{D}}_{(a+ib)}$ & $I^{\mathrm{R}}_{(a+ib)}$ \\
    \hline
    \end{tabular}
    \label{Table:OCmT}
\end{table}

\item \textit{Calculating the multi-DoF Stokes parameters and reconstructing the coherence matrix.} One may define generalized Stokes parameters to encompass two DoFs in analogy to the Stokes parameters used to characterize two-photon quantum states \cite{Abouraddy02OptComm}. In terms of the multi-DoF Stokes parameters $S_{lm}$, where $l,m\!=\!0,1,2,3$, the multi-DoF coherence matrix $\mathbf{G}$ is given by:
\begin{equation}
\mathbf{G}=\frac{1}{4}\sum_{l,m=0}^{3} S_{lm} \hat{\sigma}^{\mathrm{p}}_{l}\otimes\hat{\sigma}^{\mathrm{s}}_{m},    
\end{equation}
where $\hat{\sigma}_{l}$ and $\hat{\sigma}_{m}$ represent the standard Pauli matrices, and the super-indices s and p refer to the spatial and polarization DoFs, respectively. The coherence matrix $\mathbf{G}$ is thus given explicitly as follows:
\begin{equation}
    \mathbf{G}=
   \resizebox{.87\hsize}{!}{$
    \left(\begin{array}{cccc}
    S_{00} + S_{01} + S_{10} + S_{11} 
    & S_{02} + S_{12} - i(S_{03} + S_{13}) 
    & S_{20} + S_{21} - i(S_{30} + S_{31}) 
    & S_{22} - S_{33} - i(S_{23} + S_{32}\\
    S_{02} + S_{12} + i(S_{03} + S_{13}) 
    & S_{00} - S_{01} + S_{10} - S_{11} 
    & S_{22} - S_{33} + i(S_{23} - S_{32})
    & S_{20} - S_{21} - i(S_{30} - S_{31})\\
    S_{20} + S_{21} + i(S_{30} + S_{31})  
    & S_{22} + S_{33} - i(S_{23} - S_{32})  
    & S_{00} + S_{01} - S_{10} - S_{11} 
    & S_{02} - S_{12} - i(S_{03} - S_{13})\\
    S_{22} - S_{33} + i(S_{23} + S_{32})  
    & S_{20} - S_{21} + i(S_{30} - S_{31})  
    & S_{02} - S_{12} + i(S_{03} - S_{13})  
    & S_{00} - S_{01} - S_{10} + S_{11} \\
    \end{array}\right)$}. 
    \label{eq:G with Stokes}
    \end{equation} 

The multi-DoF Stokes parameters are defined in terms of intensity measurements as follows:
\begin{equation}
S_{lm}=4I_{lm}-2I_{0m}-2I_{l0}+I_{00},
\end{equation}
where the intensity values $I_{lm}$ are defined in terms of the measured intensity projections as follows:
\begin{eqnarray}
I_{00}&=&I^{\mathrm{H}}_{a}+I^{\mathrm{V}}_{a}+I^{\mathrm{H}}_{b}+I^{\mathrm{V}}_{b},\nonumber\\
I_{01}&=&I^{\mathrm{H}}_{a} + I^{\mathrm{H}}_{b},\nonumber\\
I_{02}&=& I^{\mathrm{D}}_{a} + I^{\mathrm{D}}_{b},\nonumber\\
I_{03}&=&I^{\mathrm{R}}_{a} + I^{\mathrm{R}}_{b},\nonumber\\
I_{10}&=&I^{\mathrm{H}}_{a} + I^{\mathrm{V}}_{a},\nonumber\\
I_{11}&=&I^{\mathrm{H}}_{a},\nonumber\\
I_{12}&=&I^{\mathrm{D}}_{a},\nonumber\\
I_{13}&=&I^{\mathrm{R}}_{a},\nonumber\\
I_{20}&=&\tfrac{1}{2}\left(I^{\mathrm{H}}_{(a+b)}+I^{\mathrm{V}}_{(a+b)}\right),\nonumber\\
I_{21}&=&\tfrac{1}{2}I^{\mathrm{H}}_{(a+b)},\nonumber\\
I_{22}&=&\tfrac{1}{2}I^{\mathrm{D}}_{(a+b)},\nonumber\\
I_{23}&=&\tfrac{1}{2}I^{\mathrm{R}}_{(a+b)},\nonumber\\
I_{30}&=&\tfrac{1}{2}\left(I^{\mathrm{H}}_{(a+ib)}+I^{\mathrm{V}}_{(a+ib)}\right),\nonumber\\
I_{31}&=&\tfrac{1}{2}I^{\mathrm{H}}_{(a+ib)},\nonumber\\
I_{32}&=&\tfrac{1}{2}I^{\mathrm{D}}_{(a+ib)},\nonumber\\
I_{33}&=&\tfrac{1}{2}I^{\mathrm{R}}_{(a+ib)}.
\end{eqnarray}

\end{enumerate}

Throughout, we employed a maximum likelihood estimation technique (described in Ref.~\cite{PhysRevA.64.052312}) to estimate the physically realizable matrices that satisfy the criteria for a coherence matrix (in particular, to exclude coherence matrices that violate positivity).

\section{References}
\bibliography{combined}

\begin{thebibliography}{52}%
\makeatletter
\providecommand \@ifxundefined [1]{%
 \@ifx{#1\undefined}
}%
\providecommand \@ifnum [1]{%
 \ifnum #1\expandafter \@firstoftwo
 \else \expandafter \@secondoftwo
 \fi
}%
\providecommand \@ifx [1]{%
 \ifx #1\expandafter \@firstoftwo
 \else \expandafter \@secondoftwo
 \fi
}%
\providecommand \natexlab [1]{#1}%
\providecommand \enquote  [1]{``#1''}%
\providecommand \bibnamefont  [1]{#1}%
\providecommand \bibfnamefont [1]{#1}%
\providecommand \citenamefont [1]{#1}%
\providecommand \href@noop [0]{\@secondoftwo}%
\providecommand \href [0]{\begingroup \@sanitize@url \@href}%
\providecommand \@href[1]{\@@startlink{#1}\@@href}%
\providecommand \@@href[1]{\endgroup#1\@@endlink}%
\providecommand \@sanitize@url [0]{\catcode `\\12\catcode `\$12\catcode `\&12\catcode `\#12\catcode `\^12\catcode `\_12\catcode `\%12\relax}%
\providecommand \@@startlink[1]{}%
\providecommand \@@endlink[0]{}%
\providecommand \url  [0]{\begingroup\@sanitize@url \@url }%
\providecommand \@url [1]{\endgroup\@href {#1}{\urlprefix }}%
\providecommand \urlprefix  [0]{URL }%
\providecommand \Eprint [0]{\href }%
\providecommand \doibase [0]{http://dx.doi.org/}%
\providecommand \selectlanguage [0]{\@gobble}%
\providecommand \bibinfo  [0]{\@secondoftwo}%
\providecommand \bibfield  [0]{\@secondoftwo}%
\providecommand \translation [1]{[#1]}%
\providecommand \BibitemOpen [0]{}%
\providecommand \bibitemStop [0]{}%
\providecommand \bibitemNoStop [0]{.\EOS\space}%
\providecommand \EOS [0]{\spacefactor3000\relax}%
\providecommand \BibitemShut  [1]{\csname bibitem#1\endcsname}%
\let\auto@bib@innerbib\@empty
\bibitem [{\citenamefont {Landauer}(1991)}]{Landauer91PT}%
  \BibitemOpen
  \bibfield  {author} {\bibinfo {author} {\bibfnamefont {R.}~\bibnamefont {Landauer}},\ }\bibfield  {title} {\enquote {\bibinfo {title} {Information is physical},}\ }\href@noop {} {\bibfield  {journal} {\bibinfo  {journal} {Phys. Today}\ }\textbf {\bibinfo {volume} {44}},\ \bibinfo {pages} {23--29} (\bibinfo {year} {1991})}\BibitemShut {NoStop}%
\bibitem [{\citenamefont {Agrawal}(2010)}]{Agrawal2010Book}%
  \BibitemOpen
  \bibfield  {author} {\bibinfo {author} {\bibfnamefont {G.~P.}\ \bibnamefont {Agrawal}},\ }\href@noop {} {\emph {\bibinfo {title} {Fiber-Optic Communication Systems}}}\ (\bibinfo  {publisher} {Wiley},\ \bibinfo {address} {New York},\ \bibinfo {year} {2010})\BibitemShut {NoStop}%
\bibitem [{\citenamefont {Ursin}\ \emph {et~al.}(2007)\citenamefont {Ursin}, \citenamefont {Tiefenbacher}, \citenamefont {Schmitt-Manderbach}, \citenamefont {Weier}, \citenamefont {Scheidl}, \citenamefont {Lindenthal}, \citenamefont {Blauensteiner}, \citenamefont {Jennewein}, \citenamefont {Perdigues}, \citenamefont {Trojek}, \citenamefont {Ömer}, \citenamefont {Fürst}, \citenamefont {Meyenburg}, \citenamefont {Rarity}, \citenamefont {Sodnik}, \citenamefont {Barbieri}, \citenamefont {Weinfurter},\ and\ \citenamefont {Zeilinger}}]{Ursin07NPhys}%
  \BibitemOpen
  \bibfield  {author} {\bibinfo {author} {\bibfnamefont {R.}~\bibnamefont {Ursin}}, \bibinfo {author} {\bibfnamefont {F.}~\bibnamefont {Tiefenbacher}}, \bibinfo {author} {\bibfnamefont {T.}~\bibnamefont {Schmitt-Manderbach}}, \bibinfo {author} {\bibfnamefont {H.}~\bibnamefont {Weier}}, \bibinfo {author} {\bibfnamefont {T.}~\bibnamefont {Scheidl}}, \bibinfo {author} {\bibfnamefont {M.}~\bibnamefont {Lindenthal}}, \bibinfo {author} {\bibfnamefont {B.}~\bibnamefont {Blauensteiner}}, \bibinfo {author} {\bibfnamefont {T.}~\bibnamefont {Jennewein}}, \bibinfo {author} {\bibfnamefont {J.}~\bibnamefont {Perdigues}}, \bibinfo {author} {\bibfnamefont {P.}~\bibnamefont {Trojek}}, \bibinfo {author} {\bibfnamefont {B.}~\bibnamefont {Ömer}}, \bibinfo {author} {\bibfnamefont {M.}~\bibnamefont {Fürst}}, \bibinfo {author} {\bibfnamefont {M.}~\bibnamefont {Meyenburg}}, \bibinfo {author} {\bibfnamefont {J.}~\bibnamefont {Rarity}}, \bibinfo {author} {\bibfnamefont {Z.}~\bibnamefont {Sodnik}}, \bibinfo {author} {\bibfnamefont
  {C.}~\bibnamefont {Barbieri}}, \bibinfo {author} {\bibfnamefont {H.}~\bibnamefont {Weinfurter}}, \ and\ \bibinfo {author} {\bibfnamefont {A.}~\bibnamefont {Zeilinger}},\ }\bibfield  {title} {\enquote {\bibinfo {title} {Entanglement-based quantum communication over 144 km},}\ }\href@noop {} {\bibfield  {journal} {\bibinfo  {journal} {Nat. Phys.}\ }\textbf {\bibinfo {volume} {3}},\ \bibinfo {pages} {481--486} (\bibinfo {year} {2007})}\BibitemShut {NoStop}%
\bibitem [{\citenamefont {Li}\ \emph {et~al.}(2014)\citenamefont {Li}, \citenamefont {Bai}, \citenamefont {Zhao},\ and\ \citenamefont {Xia}}]{Li14AOP}%
  \BibitemOpen
  \bibfield  {author} {\bibinfo {author} {\bibfnamefont {G.}~\bibnamefont {Li}}, \bibinfo {author} {\bibfnamefont {N.}~\bibnamefont {Bai}}, \bibinfo {author} {\bibfnamefont {N.}~\bibnamefont {Zhao}}, \ and\ \bibinfo {author} {\bibfnamefont {C.}~\bibnamefont {Xia}},\ }\bibfield  {title} {\enquote {\bibinfo {title} {Space-division multiplexing: the next frontier in optical communication},}\ }\href@noop {} {\bibfield  {journal} {\bibinfo  {journal} {Adv. Opt. Photon.}\ }\textbf {\bibinfo {volume} {6}},\ \bibinfo {pages} {413--487} (\bibinfo {year} {2014})}\BibitemShut {NoStop}%
\bibitem [{\citenamefont {Willner}\ \emph {et~al.}(2015)\citenamefont {Willner}, \citenamefont {Huang}, \citenamefont {Yan}, \citenamefont {Ren}, \citenamefont {Ahmed}, \citenamefont {Xie}, \citenamefont {Bao}, \citenamefont {Li}, \citenamefont {Cao}, \citenamefont {Zhao}, \citenamefont {Wang}, \citenamefont {Lavery}, \citenamefont {Tur}, \citenamefont {Ramachandran}, \citenamefont {Molisch}, \citenamefont {Ashrafi},\ and\ \citenamefont {Ashrafi}}]{Willner15AOP}%
  \BibitemOpen
  \bibfield  {author} {\bibinfo {author} {\bibfnamefont {A.~E.}\ \bibnamefont {Willner}}, \bibinfo {author} {\bibfnamefont {H.}~\bibnamefont {Huang}}, \bibinfo {author} {\bibfnamefont {Y.}~\bibnamefont {Yan}}, \bibinfo {author} {\bibfnamefont {Y.}~\bibnamefont {Ren}}, \bibinfo {author} {\bibfnamefont {N.}~\bibnamefont {Ahmed}}, \bibinfo {author} {\bibfnamefont {G.}~\bibnamefont {Xie}}, \bibinfo {author} {\bibfnamefont {C.}~\bibnamefont {Bao}}, \bibinfo {author} {\bibfnamefont {L.}~\bibnamefont {Li}}, \bibinfo {author} {\bibfnamefont {Y.}~\bibnamefont {Cao}}, \bibinfo {author} {\bibfnamefont {Z.}~\bibnamefont {Zhao}}, \bibinfo {author} {\bibfnamefont {J.}~\bibnamefont {Wang}}, \bibinfo {author} {\bibfnamefont {M.~P.~J.}\ \bibnamefont {Lavery}}, \bibinfo {author} {\bibfnamefont {M.}~\bibnamefont {Tur}}, \bibinfo {author} {\bibfnamefont {S.}~\bibnamefont {Ramachandran}}, \bibinfo {author} {\bibfnamefont {A.~F.}\ \bibnamefont {Molisch}}, \bibinfo {author} {\bibfnamefont {N.}~\bibnamefont {Ashrafi}}, \ and\
  \bibinfo {author} {\bibfnamefont {S.}~\bibnamefont {Ashrafi}},\ }\bibfield  {title} {\enquote {\bibinfo {title} {Optical communications using orbital angular momentum beams},}\ }\href@noop {} {\bibfield  {journal} {\bibinfo  {journal} {Adv. Opt. Photon.}\ }\textbf {\bibinfo {volume} {7}},\ \bibinfo {pages} {66--106} (\bibinfo {year} {2015})}\BibitemShut {NoStop}%
\bibitem [{\citenamefont {Ho}\ and\ \citenamefont {Kahn}(2014)}]{Ho14JLT}%
  \BibitemOpen
  \bibfield  {author} {\bibinfo {author} {\bibfnamefont {K.-P.}\ \bibnamefont {Ho}}\ and\ \bibinfo {author} {\bibfnamefont {J.~M.}\ \bibnamefont {Kahn}},\ }\bibfield  {title} {\enquote {\bibinfo {title} {Linear propagation effects in mode-division multiplexing systems},}\ }\href@noop {} {\bibfield  {journal} {\bibinfo  {journal} {J. Lightwave Technol.}\ }\textbf {\bibinfo {volume} {32}},\ \bibinfo {pages} {614--628} (\bibinfo {year} {2014})}\BibitemShut {NoStop}%
\bibitem [{\citenamefont {Cao}\ \emph {et~al.}(2023)\citenamefont {Cao}, \citenamefont {{\v C}i{\v z}m{\'a}r}, \citenamefont {Turtaev}, \citenamefont {Tyc},\ and\ \citenamefont {Rotter}}]{Cao23AOP}%
  \BibitemOpen
  \bibfield  {author} {\bibinfo {author} {\bibfnamefont {H.}~\bibnamefont {Cao}}, \bibinfo {author} {\bibfnamefont {T.}~\bibnamefont {{\v C}i{\v z}m{\'a}r}}, \bibinfo {author} {\bibfnamefont {S.}~\bibnamefont {Turtaev}}, \bibinfo {author} {\bibfnamefont {T.}~\bibnamefont {Tyc}}, \ and\ \bibinfo {author} {\bibfnamefont {S.}~\bibnamefont {Rotter}},\ }\bibfield  {title} {\enquote {\bibinfo {title} {Controlling light propagation in multimode fibers for imaging, spectroscopy, and beyond},}\ }\href@noop {} {\bibfield  {journal} {\bibinfo  {journal} {Adv. Opt. Photon.}\ }\textbf {\bibinfo {volume} {15}},\ \bibinfo {pages} {524--612} (\bibinfo {year} {2023})}\BibitemShut {NoStop}%
\bibitem [{\citenamefont {Tyson}(1991)}]{Tyson91Book}%
  \BibitemOpen
  \bibfield  {author} {\bibinfo {author} {\bibfnamefont {R.~K.}\ \bibnamefont {Tyson}},\ }\href@noop {} {\emph {\bibinfo {title} {Principles of Adaptive Optics}}}\ (\bibinfo  {publisher} {Academic Press},\ \bibinfo {address} {Boston},\ \bibinfo {year} {1991})\BibitemShut {NoStop}%
\bibitem [{\citenamefont {Duffner}\ and\ \citenamefont {Fugate}(2009)}]{Duffner09Book}%
  \BibitemOpen
  \bibfield  {author} {\bibinfo {author} {\bibfnamefont {R.~W.}\ \bibnamefont {Duffner}}\ and\ \bibinfo {author} {\bibfnamefont {R.~Q.}\ \bibnamefont {Fugate}},\ }\href@noop {} {\emph {\bibinfo {title} {The Adaptive Optics Revolution: A History}}}\ (\bibinfo  {publisher} {Univ. New Mexico Press},\ \bibinfo {year} {2009})\BibitemShut {NoStop}%
\bibitem [{\citenamefont {Ren}\ \emph {et~al.}(2014)\citenamefont {Ren}, \citenamefont {Xie}, \citenamefont {Huang}, \citenamefont {Ahmed}, \citenamefont {Yan}, \citenamefont {Li}, \citenamefont {Bao}, \citenamefont {Lavery}, \citenamefont {Tur}, \citenamefont {Neifeld}, \citenamefont {Boyd}, \citenamefont {Shapiro},\ and\ \citenamefont {Willner}}]{Ren14Optica}%
  \BibitemOpen
  \bibfield  {author} {\bibinfo {author} {\bibfnamefont {Y.}~\bibnamefont {Ren}}, \bibinfo {author} {\bibfnamefont {G.}~\bibnamefont {Xie}}, \bibinfo {author} {\bibfnamefont {H.}~\bibnamefont {Huang}}, \bibinfo {author} {\bibfnamefont {N.}~\bibnamefont {Ahmed}}, \bibinfo {author} {\bibfnamefont {Y.}~\bibnamefont {Yan}}, \bibinfo {author} {\bibfnamefont {L.}~\bibnamefont {Li}}, \bibinfo {author} {\bibfnamefont {C.}~\bibnamefont {Bao}}, \bibinfo {author} {\bibfnamefont {M.~P.~J.}\ \bibnamefont {Lavery}}, \bibinfo {author} {\bibfnamefont {M.}~\bibnamefont {Tur}}, \bibinfo {author} {\bibfnamefont {M.~A.}\ \bibnamefont {Neifeld}}, \bibinfo {author} {\bibfnamefont {R.~W.}\ \bibnamefont {Boyd}}, \bibinfo {author} {\bibfnamefont {J.~H.}\ \bibnamefont {Shapiro}}, \ and\ \bibinfo {author} {\bibfnamefont {A.~E.}\ \bibnamefont {Willner}},\ }\bibfield  {title} {\enquote {\bibinfo {title} {Adaptive-optics-based simultaneous pre- and post-turbulence compensation of multiple orbital-angular-momentum beams in a bidirectional
  free-space optical link},}\ }\href@noop {} {\bibfield  {journal} {\bibinfo  {journal} {Optica}\ }\textbf {\bibinfo {volume} {1}},\ \bibinfo {pages} {376--382} (\bibinfo {year} {2014})}\BibitemShut {NoStop}%
\bibitem [{\citenamefont {Carpenter}, \citenamefont {Eggleton},\ and\ \citenamefont {Schr{\"o}der}(2015)}]{Carpenter15NPhot}%
  \BibitemOpen
  \bibfield  {author} {\bibinfo {author} {\bibfnamefont {J.}~\bibnamefont {Carpenter}}, \bibinfo {author} {\bibfnamefont {B.~J.}\ \bibnamefont {Eggleton}}, \ and\ \bibinfo {author} {\bibfnamefont {J.}~\bibnamefont {Schr{\"o}der}},\ }\bibfield  {title} {\enquote {\bibinfo {title} {Observation of {E}isenbud--{W}igner--{S}mith states as principal modes in multimode fibre},}\ }\href@noop {} {\bibfield  {journal} {\bibinfo  {journal} {Nat. Photon.}\ }\textbf {\bibinfo {volume} {9}},\ \bibinfo {pages} {751--757} (\bibinfo {year} {2015})}\BibitemShut {NoStop}%
\bibitem [{\citenamefont {Defienne}, \citenamefont {Reichert},\ and\ \citenamefont {Fleischer}(2018)}]{Defienne18PRL}%
  \BibitemOpen
  \bibfield  {author} {\bibinfo {author} {\bibfnamefont {H.}~\bibnamefont {Defienne}}, \bibinfo {author} {\bibfnamefont {M.}~\bibnamefont {Reichert}}, \ and\ \bibinfo {author} {\bibfnamefont {J.~W.}\ \bibnamefont {Fleischer}},\ }\bibfield  {title} {\enquote {\bibinfo {title} {Adaptive quantum optics with spatially entangled photon pairs},}\ }\href@noop {} {\bibfield  {journal} {\bibinfo  {journal} {Phys. Rev. Lett.}\ }\textbf {\bibinfo {volume} {121}},\ \bibinfo {pages} {233601} (\bibinfo {year} {2018})}\BibitemShut {NoStop}%
\bibitem [{\citenamefont {Valencia}\ \emph {et~al.}(2020)\citenamefont {Valencia}, \citenamefont {Goel}, \citenamefont {McCutcheon}, \citenamefont {Defienne},\ and\ \citenamefont {Malik}}]{Valencia20NPhys}%
  \BibitemOpen
  \bibfield  {author} {\bibinfo {author} {\bibfnamefont {N.~H.}\ \bibnamefont {Valencia}}, \bibinfo {author} {\bibfnamefont {S.}~\bibnamefont {Goel}}, \bibinfo {author} {\bibfnamefont {W.}~\bibnamefont {McCutcheon}}, \bibinfo {author} {\bibfnamefont {H.}~\bibnamefont {Defienne}}, \ and\ \bibinfo {author} {\bibfnamefont {M.}~\bibnamefont {Malik}},\ }\bibfield  {title} {\enquote {\bibinfo {title} {Unscrambling entanglement through a complex medium},}\ }\href@noop {} {\bibfield  {journal} {\bibinfo  {journal} {Nat. Phys.}\ }\textbf {\bibinfo {volume} {16}},\ \bibinfo {pages} {1112--1116} (\bibinfo {year} {2020})}\BibitemShut {NoStop}%
\bibitem [{\citenamefont {Zhou}\ \emph {et~al.}(2021)\citenamefont {Zhou}, \citenamefont {Braverman}, \citenamefont {Fyffe}, \citenamefont {Zhang}, \citenamefont {Zhao}, \citenamefont {Willner}, \citenamefont {Shi},\ and\ \citenamefont {Boyd}}]{Zhou21NC}%
  \BibitemOpen
  \bibfield  {author} {\bibinfo {author} {\bibfnamefont {Y.}~\bibnamefont {Zhou}}, \bibinfo {author} {\bibfnamefont {B.}~\bibnamefont {Braverman}}, \bibinfo {author} {\bibfnamefont {A.}~\bibnamefont {Fyffe}}, \bibinfo {author} {\bibfnamefont {R.}~\bibnamefont {Zhang}}, \bibinfo {author} {\bibfnamefont {J.}~\bibnamefont {Zhao}}, \bibinfo {author} {\bibfnamefont {A.~E.}\ \bibnamefont {Willner}}, \bibinfo {author} {\bibfnamefont {Z.}~\bibnamefont {Shi}}, \ and\ \bibinfo {author} {\bibfnamefont {R.~W.}\ \bibnamefont {Boyd}},\ }\bibfield  {title} {\enquote {\bibinfo {title} {High-fidelity spatial mode transmission through a 1-km-long multimode fiber via vectorial time reversal},}\ }\href@noop {} {\bibfield  {journal} {\bibinfo  {journal} {Nat. Commun.}\ }\textbf {\bibinfo {volume} {12}},\ \bibinfo {pages} {1866} (\bibinfo {year} {2021})}\BibitemShut {NoStop}%
\bibitem [{\citenamefont {Walton}\ \emph {et~al.}(2003)\citenamefont {Walton}, \citenamefont {Abouraddy}, \citenamefont {Sergienko}, \citenamefont {Saleh},\ and\ \citenamefont {Teich}}]{Walton03PRL}%
  \BibitemOpen
  \bibfield  {author} {\bibinfo {author} {\bibfnamefont {Z.~D.}\ \bibnamefont {Walton}}, \bibinfo {author} {\bibfnamefont {A.~F.}\ \bibnamefont {Abouraddy}}, \bibinfo {author} {\bibfnamefont {A.~V.}\ \bibnamefont {Sergienko}}, \bibinfo {author} {\bibfnamefont {B.~E.~A.}\ \bibnamefont {Saleh}}, \ and\ \bibinfo {author} {\bibfnamefont {M.~C.}\ \bibnamefont {Teich}},\ }\bibfield  {title} {\enquote {\bibinfo {title} {Decoherence-free subspaces in quantum key distribution},}\ }\href@noop {} {\bibfield  {journal} {\bibinfo  {journal} {Phys. Rev. Lett.}\ }\textbf {\bibinfo {volume} {91}},\ \bibinfo {pages} {087901} (\bibinfo {year} {2003})}\BibitemShut {NoStop}%
\bibitem [{\citenamefont {Nelson}\ \emph {et~al.}(2014)\citenamefont {Nelson}, \citenamefont {Palastro}, \citenamefont {Davis},\ and\ \citenamefont {Sprangle}}]{Nelson14JOSAA}%
  \BibitemOpen
  \bibfield  {author} {\bibinfo {author} {\bibfnamefont {W.}~\bibnamefont {Nelson}}, \bibinfo {author} {\bibfnamefont {J.~P.}\ \bibnamefont {Palastro}}, \bibinfo {author} {\bibfnamefont {C.~C.}\ \bibnamefont {Davis}}, \ and\ \bibinfo {author} {\bibfnamefont {P.}~\bibnamefont {Sprangle}},\ }\bibfield  {title} {\enquote {\bibinfo {title} {Propagation of {B}essel and {A}iry beams through atmospheric turbulence},}\ }\href@noop {} {\bibfield  {journal} {\bibinfo  {journal} {J. Opt. Soc. Am. A}\ }\textbf {\bibinfo {volume} {31}},\ \bibinfo {pages} {603--609} (\bibinfo {year} {2014})}\BibitemShut {NoStop}%
\bibitem [{\citenamefont {Zhao}\ and\ \citenamefont {Wang}(2015)}]{Zhao15OL}%
  \BibitemOpen
  \bibfield  {author} {\bibinfo {author} {\bibfnamefont {Y.}~\bibnamefont {Zhao}}\ and\ \bibinfo {author} {\bibfnamefont {J.}~\bibnamefont {Wang}},\ }\bibfield  {title} {\enquote {\bibinfo {title} {High-base vector beam encoding/decoding for visible-light communications},}\ }\href@noop {} {\bibfield  {journal} {\bibinfo  {journal} {Opt. Lett.}\ }\textbf {\bibinfo {volume} {40}},\ \bibinfo {pages} {4843--4846} (\bibinfo {year} {2015})}\BibitemShut {NoStop}%
\bibitem [{\citenamefont {Cox}\ \emph {et~al.}(2016)\citenamefont {Cox}, \citenamefont {Rosales-Guzm{\'a}n}, \citenamefont {Lavery}, \citenamefont {Versfeld},\ and\ \citenamefont {Forbes}}]{Cox16OE}%
  \BibitemOpen
  \bibfield  {author} {\bibinfo {author} {\bibfnamefont {M.~A.}\ \bibnamefont {Cox}}, \bibinfo {author} {\bibfnamefont {C.}~\bibnamefont {Rosales-Guzm{\'a}n}}, \bibinfo {author} {\bibfnamefont {M.~P.~J.}\ \bibnamefont {Lavery}}, \bibinfo {author} {\bibfnamefont {D.~J.}\ \bibnamefont {Versfeld}}, \ and\ \bibinfo {author} {\bibfnamefont {A.}~\bibnamefont {Forbes}},\ }\bibfield  {title} {\enquote {\bibinfo {title} {On the resilience of scalar and vector vortex modes in turbulence},}\ }\href@noop {} {\bibfield  {journal} {\bibinfo  {journal} {Opt. Express}\ }\textbf {\bibinfo {volume} {24}},\ \bibinfo {pages} {18105--18113} (\bibinfo {year} {2016})}\BibitemShut {NoStop}%
\bibitem [{\citenamefont {Mphuthi}, \citenamefont {Botha},\ and\ \citenamefont {Forbes}(2018)}]{Mphuthi18JOSAA}%
  \BibitemOpen
  \bibfield  {author} {\bibinfo {author} {\bibfnamefont {N.}~\bibnamefont {Mphuthi}}, \bibinfo {author} {\bibfnamefont {R.}~\bibnamefont {Botha}}, \ and\ \bibinfo {author} {\bibfnamefont {A.}~\bibnamefont {Forbes}},\ }\bibfield  {title} {\enquote {\bibinfo {title} {Are {B}essel beams resilient to aberrations and turbulence?}}\ }\href@noop {} {\bibfield  {journal} {\bibinfo  {journal} {J. Opt. Soc. Am. A}\ }\textbf {\bibinfo {volume} {35}},\ \bibinfo {pages} {1021--1027} (\bibinfo {year} {2018})}\BibitemShut {NoStop}%
\bibitem [{\citenamefont {Zhu}\ \emph {et~al.}(2021)\citenamefont {Zhu}, \citenamefont {Janasik}, \citenamefont {Fyffe}, \citenamefont {Hay}, \citenamefont {Zhou}, \citenamefont {Kantor}, \citenamefont {Winder}, \citenamefont {Boyd}, \citenamefont {Leuchs},\ and\ \citenamefont {Shi}}]{Zhu21NC}%
  \BibitemOpen
  \bibfield  {author} {\bibinfo {author} {\bibfnamefont {Z.}~\bibnamefont {Zhu}}, \bibinfo {author} {\bibfnamefont {M.}~\bibnamefont {Janasik}}, \bibinfo {author} {\bibfnamefont {A.}~\bibnamefont {Fyffe}}, \bibinfo {author} {\bibfnamefont {D.}~\bibnamefont {Hay}}, \bibinfo {author} {\bibfnamefont {Y.}~\bibnamefont {Zhou}}, \bibinfo {author} {\bibfnamefont {B.}~\bibnamefont {Kantor}}, \bibinfo {author} {\bibfnamefont {T.}~\bibnamefont {Winder}}, \bibinfo {author} {\bibfnamefont {R.~W.}\ \bibnamefont {Boyd}}, \bibinfo {author} {\bibfnamefont {G.}~\bibnamefont {Leuchs}}, \ and\ \bibinfo {author} {\bibfnamefont {Z.}~\bibnamefont {Shi}},\ }\bibfield  {title} {\enquote {\bibinfo {title} {Compensation-free high-dimensional free-space optical communication using turbulence-resilient vector beams},}\ }\href@noop {} {\bibfield  {journal} {\bibinfo  {journal} {Nat. Commun.}\ }\textbf {\bibinfo {volume} {12}},\ \bibinfo {pages} {1666} (\bibinfo {year} {2021})}\BibitemShut {NoStop}%
\bibitem [{\citenamefont {Born}\ and\ \citenamefont {Wolf}(1999)}]{Born99Book}%
  \BibitemOpen
  \bibfield  {author} {\bibinfo {author} {\bibfnamefont {M.}~\bibnamefont {Born}}\ and\ \bibinfo {author} {\bibfnamefont {E.}~\bibnamefont {Wolf}},\ }\href@noop {} {\emph {\bibinfo {title} {Principles of Optics}}},\ \bibinfo {edition} {7th}\ ed.\ (\bibinfo  {publisher} {Cambridge Univ. Press},\ \bibinfo {address} {Cambridge},\ \bibinfo {year} {1999})\BibitemShut {NoStop}%
\bibitem [{\citenamefont {Wolf}(2007)}]{Wolf07Book}%
  \BibitemOpen
  \bibfield  {author} {\bibinfo {author} {\bibfnamefont {E.}~\bibnamefont {Wolf}},\ }\href@noop {} {\emph {\bibinfo {title} {Introduction to the Theory of Coherence and Polarization of Light}}}\ (\bibinfo  {publisher} {Cambridge Univ. Press},\ \bibinfo {address} {Cambridge},\ \bibinfo {year} {2007})\BibitemShut {NoStop}%
\bibitem [{\citenamefont {Goodman}(2015)}]{Goodman15Book}%
  \BibitemOpen
  \bibfield  {author} {\bibinfo {author} {\bibfnamefont {J.~W.}\ \bibnamefont {Goodman}},\ }\href@noop {} {\emph {\bibinfo {title} {Statistical Optics}}}\ (\bibinfo  {publisher} {John Wiley \& Sons},\ \bibinfo {address} {Hoboken, New Jersey},\ \bibinfo {year} {2015})\BibitemShut {NoStop}%
\bibitem [{\citenamefont {Baleine}\ and\ \citenamefont {Dogariu}(2005)}]{Baleine05PRL}%
  \BibitemOpen
  \bibfield  {author} {\bibinfo {author} {\bibfnamefont {E.}~\bibnamefont {Baleine}}\ and\ \bibinfo {author} {\bibfnamefont {A.}~\bibnamefont {Dogariu}},\ }\bibfield  {title} {\enquote {\bibinfo {title} {Variable coherence scattering microscopy},}\ }\href@noop {} {\bibfield  {journal} {\bibinfo  {journal} {Phys. Rev. Lett.}\ }\textbf {\bibinfo {volume} {95}},\ \bibinfo {pages} {193904} (\bibinfo {year} {2005})}\BibitemShut {NoStop}%
\bibitem [{\citenamefont {Goodman}(2007)}]{Goodman07Book}%
  \BibitemOpen
  \bibfield  {author} {\bibinfo {author} {\bibfnamefont {J.~W.}\ \bibnamefont {Goodman}},\ }\href@noop {} {\emph {\bibinfo {title} {Speckle Phenomena in Optics}}}\ (\bibinfo  {publisher} {Roberts and Company Publishers},\ \bibinfo {address} {Colorado},\ \bibinfo {year} {2007})\BibitemShut {NoStop}%
\bibitem [{\citenamefont {Kondakci}, \citenamefont {Abouraddy},\ and\ \citenamefont {Saleh}(2017)}]{Kondakci17SR}%
  \BibitemOpen
  \bibfield  {author} {\bibinfo {author} {\bibfnamefont {H.~E.}\ \bibnamefont {Kondakci}}, \bibinfo {author} {\bibfnamefont {A.~F.}\ \bibnamefont {Abouraddy}}, \ and\ \bibinfo {author} {\bibfnamefont {B.~E.~A.}\ \bibnamefont {Saleh}},\ }\bibfield  {title} {\enquote {\bibinfo {title} {Lattice topology dictates photon statistics},}\ }\href@noop {} {\bibfield  {journal} {\bibinfo  {journal} {Sci. Rep.}\ }\textbf {\bibinfo {volume} {7}},\ \bibinfo {pages} {8948} (\bibinfo {year} {2017})}\BibitemShut {NoStop}%
\bibitem [{\citenamefont {Han}, \citenamefont {Bender},\ and\ \citenamefont {Cao}(2023)}]{Han23PRL}%
  \BibitemOpen
  \bibfield  {author} {\bibinfo {author} {\bibfnamefont {S.}~\bibnamefont {Han}}, \bibinfo {author} {\bibfnamefont {N.}~\bibnamefont {Bender}}, \ and\ \bibinfo {author} {\bibfnamefont {H.}~\bibnamefont {Cao}},\ }\bibfield  {title} {\enquote {\bibinfo {title} {Tailoring {3D} speckle statistics},}\ }\href@noop {} {\bibfield  {journal} {\bibinfo  {journal} {Phys. Rev. Lett.}\ }\textbf {\bibinfo {volume} {130}},\ \bibinfo {pages} {093802} (\bibinfo {year} {2023})}\BibitemShut {NoStop}%
\bibitem [{\citenamefont {Bender}\ \emph {et~al.}(2023)\citenamefont {Bender}, \citenamefont {Haig}, \citenamefont {Christodoulides},\ and\ \citenamefont {Wise}}]{Bender23Optica}%
  \BibitemOpen
  \bibfield  {author} {\bibinfo {author} {\bibfnamefont {N.}~\bibnamefont {Bender}}, \bibinfo {author} {\bibfnamefont {H.}~\bibnamefont {Haig}}, \bibinfo {author} {\bibfnamefont {D.~N.}\ \bibnamefont {Christodoulides}}, \ and\ \bibinfo {author} {\bibfnamefont {F.~W.}\ \bibnamefont {Wise}},\ }\bibfield  {title} {\enquote {\bibinfo {title} {Spectral speckle customization},}\ }\href@noop {} {\bibfield  {journal} {\bibinfo  {journal} {Optica}\ }\textbf {\bibinfo {volume} {10}},\ \bibinfo {pages} {1260--1268} (\bibinfo {year} {2023})}\BibitemShut {NoStop}%
\bibitem [{\citenamefont {Waller}, \citenamefont {Situ},\ and\ \citenamefont {Fleischer}(2012)}]{Waller12NP}%
  \BibitemOpen
  \bibfield  {author} {\bibinfo {author} {\bibfnamefont {L.}~\bibnamefont {Waller}}, \bibinfo {author} {\bibfnamefont {G.}~\bibnamefont {Situ}}, \ and\ \bibinfo {author} {\bibfnamefont {J.~W.}\ \bibnamefont {Fleischer}},\ }\bibfield  {title} {\enquote {\bibinfo {title} {Phase-space measurement and coherence synthesis of optical beams},}\ }\href@noop {} {\bibfield  {journal} {\bibinfo  {journal} {Nat. Photon.}\ }\textbf {\bibinfo {volume} {6}},\ \bibinfo {pages} {474--479} (\bibinfo {year} {2012})}\BibitemShut {NoStop}%
\bibitem [{\citenamefont {Nardi}\ \emph {et~al.}(2022)\citenamefont {Nardi}, \citenamefont {Divitt}, \citenamefont {Rossi}, \citenamefont {Tebbenjohanns}, \citenamefont {Militaru}, \citenamefont {Frimmer},\ and\ \citenamefont {Novotny}}]{Nardi22OL}%
  \BibitemOpen
  \bibfield  {author} {\bibinfo {author} {\bibfnamefont {A.}~\bibnamefont {Nardi}}, \bibinfo {author} {\bibfnamefont {S.}~\bibnamefont {Divitt}}, \bibinfo {author} {\bibfnamefont {M.}~\bibnamefont {Rossi}}, \bibinfo {author} {\bibfnamefont {F.}~\bibnamefont {Tebbenjohanns}}, \bibinfo {author} {\bibfnamefont {A.}~\bibnamefont {Militaru}}, \bibinfo {author} {\bibfnamefont {M.}~\bibnamefont {Frimmer}}, \ and\ \bibinfo {author} {\bibfnamefont {L.}~\bibnamefont {Novotny}},\ }\bibfield  {title} {\enquote {\bibinfo {title} {Encoding information in the mutual coherence of spatially separated light beams},}\ }\href@noop {} {\bibfield  {journal} {\bibinfo  {journal} {Opt. Lett.}\ }\textbf {\bibinfo {volume} {47}},\ \bibinfo {pages} {4588--4591} (\bibinfo {year} {2022})}\BibitemShut {NoStop}%
\bibitem [{\citenamefont {Vitullo}\ \emph {et~al.}(2017)\citenamefont {Vitullo}, \citenamefont {Leary}, \citenamefont {Gregg}, \citenamefont {Smith}, \citenamefont {Reddy}, \citenamefont {Ramachandran},\ and\ \citenamefont {Raymer}}]{Vitullo17PRL}%
  \BibitemOpen
  \bibfield  {author} {\bibinfo {author} {\bibfnamefont {D.~L.~P.}\ \bibnamefont {Vitullo}}, \bibinfo {author} {\bibfnamefont {C.~C.}\ \bibnamefont {Leary}}, \bibinfo {author} {\bibfnamefont {P.}~\bibnamefont {Gregg}}, \bibinfo {author} {\bibfnamefont {R.~A.}\ \bibnamefont {Smith}}, \bibinfo {author} {\bibfnamefont {D.~V.}\ \bibnamefont {Reddy}}, \bibinfo {author} {\bibfnamefont {S.}~\bibnamefont {Ramachandran}}, \ and\ \bibinfo {author} {\bibfnamefont {M.~G.}\ \bibnamefont {Raymer}},\ }\bibfield  {title} {\enquote {\bibinfo {title} {Observation of interaction of spin and intrinsic orbital angular momentum of light},}\ }\href@noop {} {\bibfield  {journal} {\bibinfo  {journal} {Phys. Rev. Lett.}\ }\textbf {\bibinfo {volume} {118}},\ \bibinfo {pages} {083601} (\bibinfo {year} {2017})}\BibitemShut {NoStop}%
\bibitem [{\citenamefont {Xiong}\ \emph {et~al.}(2018)\citenamefont {Xiong}, \citenamefont {Hsu}, \citenamefont {Bromberg}, \citenamefont {Antonio-Lopez}, \citenamefont {{Amezcua Correa}},\ and\ \citenamefont {Cao}}]{Xiong2018}%
  \BibitemOpen
  \bibfield  {author} {\bibinfo {author} {\bibfnamefont {W.}~\bibnamefont {Xiong}}, \bibinfo {author} {\bibfnamefont {C.}~\bibnamefont {Hsu}}, \bibinfo {author} {\bibfnamefont {Y.}~\bibnamefont {Bromberg}}, \bibinfo {author} {\bibfnamefont {J.}~\bibnamefont {Antonio-Lopez}}, \bibinfo {author} {\bibfnamefont {R.}~\bibnamefont {{Amezcua Correa}}}, \ and\ \bibinfo {author} {\bibfnamefont {H.}~\bibnamefont {Cao}},\ }\bibfield  {title} {\enquote {\bibinfo {title} {{Complete polarization control in multimode fibers with polarization and mode coupling}},}\ }\href@noop {} {\bibfield  {journal} {\bibinfo  {journal} {Light Sci. Appl.}\ }\textbf {\bibinfo {volume} {7}},\ \bibinfo {pages} {1--10} (\bibinfo {year} {2018})}\BibitemShut {NoStop}%
\bibitem [{\citenamefont {Abouraddy}(2017)}]{Abouraddy17OE}%
  \BibitemOpen
  \bibfield  {author} {\bibinfo {author} {\bibfnamefont {A.~F.}\ \bibnamefont {Abouraddy}},\ }\bibfield  {title} {\enquote {\bibinfo {title} {What is the maximum attainable visibility by a partially coherent electromagnetic field in {Y}oung's double-slit interference?}}\ }\href@noop {} {\bibfield  {journal} {\bibinfo  {journal} {Opt. Express}\ }\textbf {\bibinfo {volume} {25}},\ \bibinfo {pages} {18320--18331} (\bibinfo {year} {2017})}\BibitemShut {NoStop}%
\bibitem [{\citenamefont {Okoro}\ \emph {et~al.}(2017)\citenamefont {Okoro}, \citenamefont {Kondakci}, \citenamefont {Abouraddy},\ and\ \citenamefont {Toussaint}}]{Okoro17Optica}%
  \BibitemOpen
  \bibfield  {author} {\bibinfo {author} {\bibfnamefont {C.}~\bibnamefont {Okoro}}, \bibinfo {author} {\bibfnamefont {H.~E.}\ \bibnamefont {Kondakci}}, \bibinfo {author} {\bibfnamefont {A.~F.}\ \bibnamefont {Abouraddy}}, \ and\ \bibinfo {author} {\bibfnamefont {K.~C.}\ \bibnamefont {Toussaint}},\ }\bibfield  {title} {\enquote {\bibinfo {title} {Demonstration of an optical-coherence converter},}\ }\href@noop {} {\bibfield  {journal} {\bibinfo  {journal} {Optica}\ }\textbf {\bibinfo {volume} {4}},\ \bibinfo {pages} {1052--1058} (\bibinfo {year} {2017})}\BibitemShut {NoStop}%
\bibitem [{\citenamefont {Harling}\ \emph {et~al.}(2022)\citenamefont {Harling}, \citenamefont {Kelkar}, \citenamefont {Okoro}, \citenamefont {Diouf}, \citenamefont {Abouraddy},\ and\ \citenamefont {Toussaint}}]{harling2022reversible}%
  \BibitemOpen
  \bibfield  {author} {\bibinfo {author} {\bibfnamefont {M.}~\bibnamefont {Harling}}, \bibinfo {author} {\bibfnamefont {V.}~\bibnamefont {Kelkar}}, \bibinfo {author} {\bibfnamefont {C.}~\bibnamefont {Okoro}}, \bibinfo {author} {\bibfnamefont {M.}~\bibnamefont {Diouf}}, \bibinfo {author} {\bibfnamefont {A.~F.}\ \bibnamefont {Abouraddy}}, \ and\ \bibinfo {author} {\bibfnamefont {K.~C.}\ \bibnamefont {Toussaint}},\ }\bibfield  {title} {\enquote {\bibinfo {title} {Reversible inter-degree-of-freedom optical-coherence conversion via entropy swapping},}\ }\href@noop {} {\bibfield  {journal} {\bibinfo  {journal} {Opt. Express}\ }\textbf {\bibinfo {volume} {30}},\ \bibinfo {pages} {29584--29597} (\bibinfo {year} {2022})}\BibitemShut {NoStop}%
\bibitem [{\citenamefont {Harling}\ \emph {et~al.}(2023)\citenamefont {Harling}, \citenamefont {Kelkar}, \citenamefont {Abouraddy},\ and\ \citenamefont {Toussaint}}]{Harling23JO}%
  \BibitemOpen
  \bibfield  {author} {\bibinfo {author} {\bibfnamefont {M.}~\bibnamefont {Harling}}, \bibinfo {author} {\bibfnamefont {V.}~\bibnamefont {Kelkar}}, \bibinfo {author} {\bibfnamefont {A.~F.}\ \bibnamefont {Abouraddy}}, \ and\ \bibinfo {author} {\bibfnamefont {K.~C.}\ \bibnamefont {Toussaint}},\ }\bibfield  {title} {\enquote {\bibinfo {title} {Reversible coherence conversion across optical degrees-of-freedom: a tutorial},}\ }\href@noop {} {\bibfield  {journal} {\bibinfo  {journal} {J. Opt.}\ }\textbf {\bibinfo {volume} {25}},\ \bibinfo {pages} {053502} (\bibinfo {year} {2023})}\BibitemShut {NoStop}%
\bibitem [{\citenamefont {Harling}\ \emph {et~al.}(2024{\natexlab{a}})\citenamefont {Harling}, \citenamefont {Kelkar}, \citenamefont {Toussaint},\ and\ \citenamefont {Abouraddy}}]{Harling24PRA}%
  \BibitemOpen
  \bibfield  {author} {\bibinfo {author} {\bibfnamefont {M.}~\bibnamefont {Harling}}, \bibinfo {author} {\bibfnamefont {V.~A.}\ \bibnamefont {Kelkar}}, \bibinfo {author} {\bibfnamefont {K.~C.}\ \bibnamefont {Toussaint}}, \ and\ \bibinfo {author} {\bibfnamefont {A.~F.}\ \bibnamefont {Abouraddy}},\ }\bibfield  {title} {\enquote {\bibinfo {title} {Locked entropy in partially coherent optical fields},}\ }\href@noop {} {\bibfield  {journal} {\bibinfo  {journal} {Phys. Rev. A}\ }\textbf {\bibinfo {volume} {109}},\ \bibinfo {pages} {L021501} (\bibinfo {year} {2024}{\natexlab{a}})}\BibitemShut {NoStop}%
\bibitem [{\citenamefont {Harling}\ \emph {et~al.}(2024{\natexlab{b}})\citenamefont {Harling}, \citenamefont {Kelkar}, \citenamefont {Toussaint},\ and\ \citenamefont {Abouraddy}}]{Harling24PRA2}%
  \BibitemOpen
  \bibfield  {author} {\bibinfo {author} {\bibfnamefont {M.}~\bibnamefont {Harling}}, \bibinfo {author} {\bibfnamefont {V.~A.}\ \bibnamefont {Kelkar}}, \bibinfo {author} {\bibfnamefont {K.~C.}\ \bibnamefont {Toussaint}}, \ and\ \bibinfo {author} {\bibfnamefont {A.~F.}\ \bibnamefont {Abouraddy}},\ }\bibfield  {title} {\enquote {\bibinfo {title} {Isoentropic partially coherent optical fields that cannot be interconverted unitarily},}\ }\href@noop {} {\bibfield  {journal} {\bibinfo  {journal} {Phys. Rev. A}\ }\textbf {\bibinfo {volume} {110}},\ \bibinfo {pages} {013505} (\bibinfo {year} {2024}{\natexlab{b}})}\BibitemShut {NoStop}%
\bibitem [{\citenamefont {Halder}, \citenamefont {Norrman},\ and\ \citenamefont {Friberg}(2021)}]{Halder21OL}%
  \BibitemOpen
  \bibfield  {author} {\bibinfo {author} {\bibfnamefont {A.}~\bibnamefont {Halder}}, \bibinfo {author} {\bibfnamefont {A.}~\bibnamefont {Norrman}}, \ and\ \bibinfo {author} {\bibfnamefont {A.~T.}\ \bibnamefont {Friberg}},\ }\bibfield  {title} {\enquote {\bibinfo {title} {Poincar{\'e} sphere representation of scalar two-beam interference under spatial unitary transformations},}\ }\href@noop {} {\bibfield  {journal} {\bibinfo  {journal} {Opt. Lett.}\ }\textbf {\bibinfo {volume} {46}},\ \bibinfo {pages} {5619--5622} (\bibinfo {year} {2021})}\BibitemShut {NoStop}%
\bibitem [{\citenamefont {Brosseau}\ and\ \citenamefont {Dogariu}(2006)}]{Brosseau06PO}%
  \BibitemOpen
  \bibfield  {author} {\bibinfo {author} {\bibfnamefont {C.}~\bibnamefont {Brosseau}}\ and\ \bibinfo {author} {\bibfnamefont {A.}~\bibnamefont {Dogariu}},\ }\bibfield  {title} {\enquote {\bibinfo {title} {Symmetry properties and polarization descriptors for an arbitrary electromagnetic wavefield},}\ }\href@noop {} {\bibfield  {journal} {\bibinfo  {journal} {Prog. Opt.}\ }\textbf {\bibinfo {volume} {49}},\ \bibinfo {pages} {315--380} (\bibinfo {year} {2006})}\BibitemShut {NoStop}%
\bibitem [{\citenamefont {Gori}, \citenamefont {Santarsiero},\ and\ \citenamefont {Borghi}(2006)}]{Gori06OL}%
  \BibitemOpen
  \bibfield  {author} {\bibinfo {author} {\bibfnamefont {F.}~\bibnamefont {Gori}}, \bibinfo {author} {\bibfnamefont {M.}~\bibnamefont {Santarsiero}}, \ and\ \bibinfo {author} {\bibfnamefont {R.}~\bibnamefont {Borghi}},\ }\bibfield  {title} {\enquote {\bibinfo {title} {Vector mode analysis of a {Y}oung interferometer},}\ }\href@noop {} {\bibfield  {journal} {\bibinfo  {journal} {Opt. Lett.}\ }\textbf {\bibinfo {volume} {31}},\ \bibinfo {pages} {858--860} (\bibinfo {year} {2006})}\BibitemShut {NoStop}%
\bibitem [{\citenamefont {Kagalwala}\ \emph {et~al.}(2013)\citenamefont {Kagalwala}, \citenamefont {{Di G}iuseppe}, \citenamefont {Abouraddy},\ and\ \citenamefont {Saleh}}]{Kagalwala13NP}%
  \BibitemOpen
  \bibfield  {author} {\bibinfo {author} {\bibfnamefont {K.~H.}\ \bibnamefont {Kagalwala}}, \bibinfo {author} {\bibfnamefont {G.}~\bibnamefont {{Di G}iuseppe}}, \bibinfo {author} {\bibfnamefont {A.~F.}\ \bibnamefont {Abouraddy}}, \ and\ \bibinfo {author} {\bibfnamefont {B.~E.~A.}\ \bibnamefont {Saleh}},\ }\bibfield  {title} {\enquote {\bibinfo {title} {Bell's measure in classical optical coherence},}\ }\href@noop {} {\bibfield  {journal} {\bibinfo  {journal} {Nat. Photon.}\ }\textbf {\bibinfo {volume} {7}},\ \bibinfo {pages} {72--78} (\bibinfo {year} {2013})}\BibitemShut {NoStop}%
\bibitem [{\citenamefont {Wolf}(1982)}]{Wolf82JOSA}%
  \BibitemOpen
  \bibfield  {author} {\bibinfo {author} {\bibfnamefont {E.}~\bibnamefont {Wolf}},\ }\bibfield  {title} {\enquote {\bibinfo {title} {New theory of partial coherence in the space–frequency domain. part {I}: spectra and cross spectra of steady-state sources},}\ }\href@noop {} {\bibfield  {journal} {\bibinfo  {journal} {J. Opt. Soc. Am.}\ }\textbf {\bibinfo {volume} {72}},\ \bibinfo {pages} {343--351} (\bibinfo {year} {1982})}\BibitemShut {NoStop}%
\bibitem [{\citenamefont {Abouraddy}, \citenamefont {Kagalwala},\ and\ \citenamefont {Saleh}(2014)}]{Abouraddy14OL}%
  \BibitemOpen
  \bibfield  {author} {\bibinfo {author} {\bibfnamefont {A.~F.}\ \bibnamefont {Abouraddy}}, \bibinfo {author} {\bibfnamefont {K.~H.}\ \bibnamefont {Kagalwala}}, \ and\ \bibinfo {author} {\bibfnamefont {B.~E.~A.}\ \bibnamefont {Saleh}},\ }\bibfield  {title} {\enquote {\bibinfo {title} {Two-point optical coherency matrix tomography},}\ }\href@noop {} {\bibfield  {journal} {\bibinfo  {journal} {Opt. Lett.}\ }\textbf {\bibinfo {volume} {39}},\ \bibinfo {pages} {2411--2414} (\bibinfo {year} {2014})}\BibitemShut {NoStop}%
\bibitem [{\citenamefont {Kagalwala}\ \emph {et~al.}(2015)\citenamefont {Kagalwala}, \citenamefont {Kondakci}, \citenamefont {Abouraddy},\ and\ \citenamefont {Saleh}}]{Kagalwala15SR}%
  \BibitemOpen
  \bibfield  {author} {\bibinfo {author} {\bibfnamefont {K.~H.}\ \bibnamefont {Kagalwala}}, \bibinfo {author} {\bibfnamefont {H.~E.}\ \bibnamefont {Kondakci}}, \bibinfo {author} {\bibfnamefont {A.~F.}\ \bibnamefont {Abouraddy}}, \ and\ \bibinfo {author} {\bibfnamefont {B.~E.~A.}\ \bibnamefont {Saleh}},\ }\bibfield  {title} {\enquote {\bibinfo {title} {Optical coherency matrix tomography},}\ }\href@noop {} {\bibfield  {journal} {\bibinfo  {journal} {Sci. Rep.}\ }\textbf {\bibinfo {volume} {5}},\ \bibinfo {pages} {15333} (\bibinfo {year} {2015})}\BibitemShut {NoStop}%
\bibitem [{\citenamefont {Dimitrov}\ and\ \citenamefont {Haas}(2015)}]{Dimitrov15Book}%
  \BibitemOpen
  \bibfield  {author} {\bibinfo {author} {\bibfnamefont {S.}~\bibnamefont {Dimitrov}}\ and\ \bibinfo {author} {\bibfnamefont {H.}~\bibnamefont {Haas}},\ }\href@noop {} {\emph {\bibinfo {title} {Principles of LED Light Communications: Towards Networked Li-Fi}}}\ (\bibinfo  {publisher} {Cambridge Univ. Press},\ \bibinfo {address} {Cambridge},\ \bibinfo {year} {2015})\BibitemShut {NoStop}%
\bibitem [{\citenamefont {James}\ \emph {et~al.}(2001{\natexlab{a}})\citenamefont {James}, \citenamefont {Kwiat}, \citenamefont {Munro},\ and\ \citenamefont {White}}]{James01PRA}%
  \BibitemOpen
  \bibfield  {author} {\bibinfo {author} {\bibfnamefont {D.~F.~V.}\ \bibnamefont {James}}, \bibinfo {author} {\bibfnamefont {P.~G.}\ \bibnamefont {Kwiat}}, \bibinfo {author} {\bibfnamefont {W.~J.}\ \bibnamefont {Munro}}, \ and\ \bibinfo {author} {\bibfnamefont {A.~G.}\ \bibnamefont {White}},\ }\bibfield  {title} {\enquote {\bibinfo {title} {Measurement of qubits},}\ }\href@noop {} {\bibfield  {journal} {\bibinfo  {journal} {Phys. Rev. A}\ }\textbf {\bibinfo {volume} {64}},\ \bibinfo {pages} {052312} (\bibinfo {year} {2001}{\natexlab{a}})}\BibitemShut {NoStop}%
\bibitem [{\citenamefont {Saleh}\ and\ \citenamefont {Teich}(2007)}]{SalehBook07}%
  \BibitemOpen
  \bibfield  {author} {\bibinfo {author} {\bibfnamefont {B.~E.~A.}\ \bibnamefont {Saleh}}\ and\ \bibinfo {author} {\bibfnamefont {M.~C.}\ \bibnamefont {Teich}},\ }\href@noop {} {\emph {\bibinfo {title} {Principles of Photonics}}}\ (\bibinfo  {publisher} {Wiley},\ \bibinfo {year} {2007})\BibitemShut {NoStop}%
\bibitem [{\citenamefont {Roques-Carmes}, \citenamefont {Fan},\ and\ \citenamefont {Miller}(2024)}]{roques2024measuring}%
  \BibitemOpen
  \bibfield  {author} {\bibinfo {author} {\bibfnamefont {C.}~\bibnamefont {Roques-Carmes}}, \bibinfo {author} {\bibfnamefont {S.}~\bibnamefont {Fan}}, \ and\ \bibinfo {author} {\bibfnamefont {D.}~\bibnamefont {Miller}},\ }\bibfield  {title} {\enquote {\bibinfo {title} {Measuring, processing, and generating partially coherent light with self-configuring optics},}\ }\href@noop {} {\bibfield  {journal} {\bibinfo  {journal} {arXiv:2402.00704}\ } (\bibinfo {year} {2024})}\BibitemShut {NoStop}%
\bibitem [{\citenamefont {Bogaerts}\ \emph {et~al.}(2020)\citenamefont {Bogaerts}, \citenamefont {P{\'e}rez}, \citenamefont {Capmany}, \citenamefont {Miller}, \citenamefont {Poon}, \citenamefont {Englund}, \citenamefont {Morichetti},\ and\ \citenamefont {Melloni}}]{Bogaerts20Nature}%
  \BibitemOpen
  \bibfield  {author} {\bibinfo {author} {\bibfnamefont {W.}~\bibnamefont {Bogaerts}}, \bibinfo {author} {\bibfnamefont {D.}~\bibnamefont {P{\'e}rez}}, \bibinfo {author} {\bibfnamefont {J.}~\bibnamefont {Capmany}}, \bibinfo {author} {\bibfnamefont {D.~A.~B.}\ \bibnamefont {Miller}}, \bibinfo {author} {\bibfnamefont {J.}~\bibnamefont {Poon}}, \bibinfo {author} {\bibfnamefont {D.}~\bibnamefont {Englund}}, \bibinfo {author} {\bibfnamefont {F.}~\bibnamefont {Morichetti}}, \ and\ \bibinfo {author} {\bibfnamefont {A.}~\bibnamefont {Melloni}},\ }\bibfield  {title} {\enquote {\bibinfo {title} {Programmable photonic circuits},}\ }\href@noop {} {\bibfield  {journal} {\bibinfo  {journal} {Nature}\ }\textbf {\bibinfo {volume} {586}},\ \bibinfo {pages} {207–216} (\bibinfo {year} {2020})}\BibitemShut {NoStop}%
\bibitem [{\citenamefont {Abouraddy}\ \emph {et~al.}(2002)\citenamefont {Abouraddy}, \citenamefont {Sergienko}, \citenamefont {Saleh},\ and\ \citenamefont {Teich}}]{Abouraddy02OptComm}%
  \BibitemOpen
  \bibfield  {author} {\bibinfo {author} {\bibfnamefont {A.~F.}\ \bibnamefont {Abouraddy}}, \bibinfo {author} {\bibfnamefont {A.~V.}\ \bibnamefont {Sergienko}}, \bibinfo {author} {\bibfnamefont {B.~E.~A.}\ \bibnamefont {Saleh}}, \ and\ \bibinfo {author} {\bibfnamefont {M.~C.}\ \bibnamefont {Teich}},\ }\bibfield  {title} {\enquote {\bibinfo {title} {Quantum entanglement and the two-photon {S}tokes parameters},}\ }\href@noop {} {\bibfield  {journal} {\bibinfo  {journal} {Opt. Comm.}\ }\textbf {\bibinfo {volume} {201}},\ \bibinfo {pages} {93--98} (\bibinfo {year} {2002})}\BibitemShut {NoStop}%
\bibitem [{\citenamefont {James}\ \emph {et~al.}(2001{\natexlab{b}})\citenamefont {James}, \citenamefont {Kwiat}, \citenamefont {Munro},\ and\ \citenamefont {White}}]{PhysRevA.64.052312}%
  \BibitemOpen
  \bibfield  {author} {\bibinfo {author} {\bibfnamefont {D.~F.~V.}\ \bibnamefont {James}}, \bibinfo {author} {\bibfnamefont {P.~G.}\ \bibnamefont {Kwiat}}, \bibinfo {author} {\bibfnamefont {W.~J.}\ \bibnamefont {Munro}}, \ and\ \bibinfo {author} {\bibfnamefont {A.~G.}\ \bibnamefont {White}},\ }\bibfield  {title} {\enquote {\bibinfo {title} {Measurement of qubits},}\ }\href {\doibase 10.1103/PhysRevA.64.052312} {\bibfield  {journal} {\bibinfo  {journal} {Phys. Rev. A}\ }\textbf {\bibinfo {volume} {64}},\ \bibinfo {pages} {052312} (\bibinfo {year} {2001}{\natexlab{b}})}\BibitemShut {NoStop}%
\end{thebibliography}%

\end{document}